\pdfoutput=1
\documentclass[useAMS,usenatbib]{mn2e}
\usepackage{xcolor}
\usepackage{amsmath, float}
\usepackage{txfonts}
\usepackage{fleqn}   
\usepackage{graphicx}
\usepackage{epstopdf}
\usepackage{subfigure}
\usepackage{verbatim}
\usepackage{mathtools}
\usepackage{bm}

\renewcommand{\thesubfigure}{(\bf{\thefigure.\arabic{subfigure}})}
\makeatletter
\renewcommand{\@thesubfigure}{\thesubfigure \space}
\renewcommand{\p@subfigure}{}
\makeatother

\DeclarePairedDelimiter\abs{\lvert}{\rvert}

\title[3D R-MHD simulations of  BSPWNe: dynamics]
{Full-3D relativistic MHD simulations of  Bow Shock Pulsar Wind Nebulae: dynamics}
\author[B. Olmi \& N. Bucciantini]{
 B. Olmi$^{1,2,3}$ \thanks{E-mail: barbara.olmi@unifi.it} \& N. Bucciantini$^{2,1,3}$
\\
$^{1}$Dipartimento di Fisica e Astronomia, Universit\`a degli Studi di Firenze, Via G. Sansone 1, 
I-50019 Sesto F.~no  (Firenze), Italy\\
$^{2}$INAF - Osservatorio Astrofisico di Arcetri, Largo E. Fermi 5,
I-50125 Firenze, Italy\\
$^{3}$INFN - Sezione di Firenze, Via G. Sansone 1, I-50019 Sesto F.~no  (Firenze), Italy\\}

\begin{document}
 
\date{Accepted / Received}

\maketitle

\label{firstpage}

\begin{abstract}
Bow shock  pulsar wind nebulae (BSPWNe) are know to show a large variety of shapes and morphologies, both when comparing different objects, and for the same object in different energy bands. It is unclear if such a variety  is related to differences in the pulsar wind properties, or to differences in the conditions of the ambient medium. We present here a set of full three-dimensional, relativistic and magneto-hydrodynamic simulations of BSPWNe, with the intention of determining how differences in the injection conditions by the pulsar wind reflect in the nebular dynamics. To achieve a good coverage of the available parameter space we have run several simulations varying those parameters that are most representative of the wind properties: the latitudinal anisotropy of the wind energy flux with respect to the pulsar spin axis, the level of magnetization, the inclination of the pulsar spin axis with respect to the pulsar direction of motion. We have followed the dynamics in these systems, not just in the very head, but also in the tail, trying to assess if and how the system retains memory of the injection at large distances from the pulsar itself. In this paper we focus our attention on the characterization of the fluid structure and magnetic field properties. We have tried to evaluate the level of turbulence in the tail, and its relation to injection, the survival of current sheets, and the degree of mixing between the shocked ambient medium and the relativistic pulsar wind material.
\end{abstract}


\section{Introduction}
\label{sec:intro}
Pulsar Wind Nebulae (PWNe) are synchrotron emitting bubbles powered by a pulsar wind. They are commonly found inside the remnant of their parent supernova explosions, but in older systems they can also arise due to the direct interaction with the ISM. The pulsar wind is a magnetized, cold and ultra-relativistic outflow, with predicted Lorentz factors in the range $10^4-10^7$  \citep{Goldreich:1969,Kennel:1984,Kennel:1984a}), and it is thought to be mainly composed by electron-positron pairs  \citep{Ruderman_Sutherland75a,Arons_Scharlemann79a, Contopoulos_Kazanas+99a,Spitkovsky06a,Tchekhovskoy_Philippov+16a,Hibschman:2001,Takata_Wang+10a,Timokhin_Arons13a,Takata_Ng+16a}. 
It is launched at the pulsar magnetosphere (at the typical distances of the light cylinder) at the expenses of the stellar rotational energy.  As a consequence of the interaction with the ambient medium, the supersonic wind is forced to slow down, and it does so by a strong termination shock (TS), where particles are likely accelerated to a non-thermal distribution. The observed radiation is produced via non-thermal processes (synchrotron and inverse Compton scattering) arising from the interplay of these particles with the nebular magnetic field and the background photon field.

Exemplary PWNe involving young pulsars inside their supernova remnant (SNR)  are the Crab and Vela nebulae. Systems like these are referred as  \textit{plerions} \citep{Gaensler_Slane06a,Bucciantini08b,Olmi:2016}.
However, given that a relevant fraction of all the pulsars, between 10\% and 50\%, is born with high kick-velocity, of the order of 100-500 km s$^{-1}$,  \citep{Cordes_Chernoff98a,Arzoumanian:2002,Sartore_Ripamonti+10a,Verbunt_Igoshev+17a}), while on the other hand the remnant expansion is decelerated \citep{Truelove_McKee99a,Cioffi_McKee+88a,Leahy_Green+14a,Sanchez-Cruces_Rosado+18a}, they are fated to escape their progenitor SNR over timescales of a few tens of thousands of years, much shorter than the pulsars typical ages of the order of $10^6$ years. 
Once outside, given the typical sound speeds in the ISM of the order of $10-100$  km s$^{-1}$, these pulsars turn out to move at a supersonic speed. No longer observed as plerion-like bubbles, their associated nebulae acquire a cometary like shape due to the balance of the ram pressure of the pulsar wind with the surrounding ISM \citep{Wilkin:1996,Bucciantini:2001,Bucciantini:2002}. The pulsar is now found  at the head of an elongated tail that extends in the direction opposite to the pulsar motion.  These objects are then known as bow shock PWNe (BSPWNe).
\begin{figure}
	\centering
	\includegraphics[width=.5\textwidth]{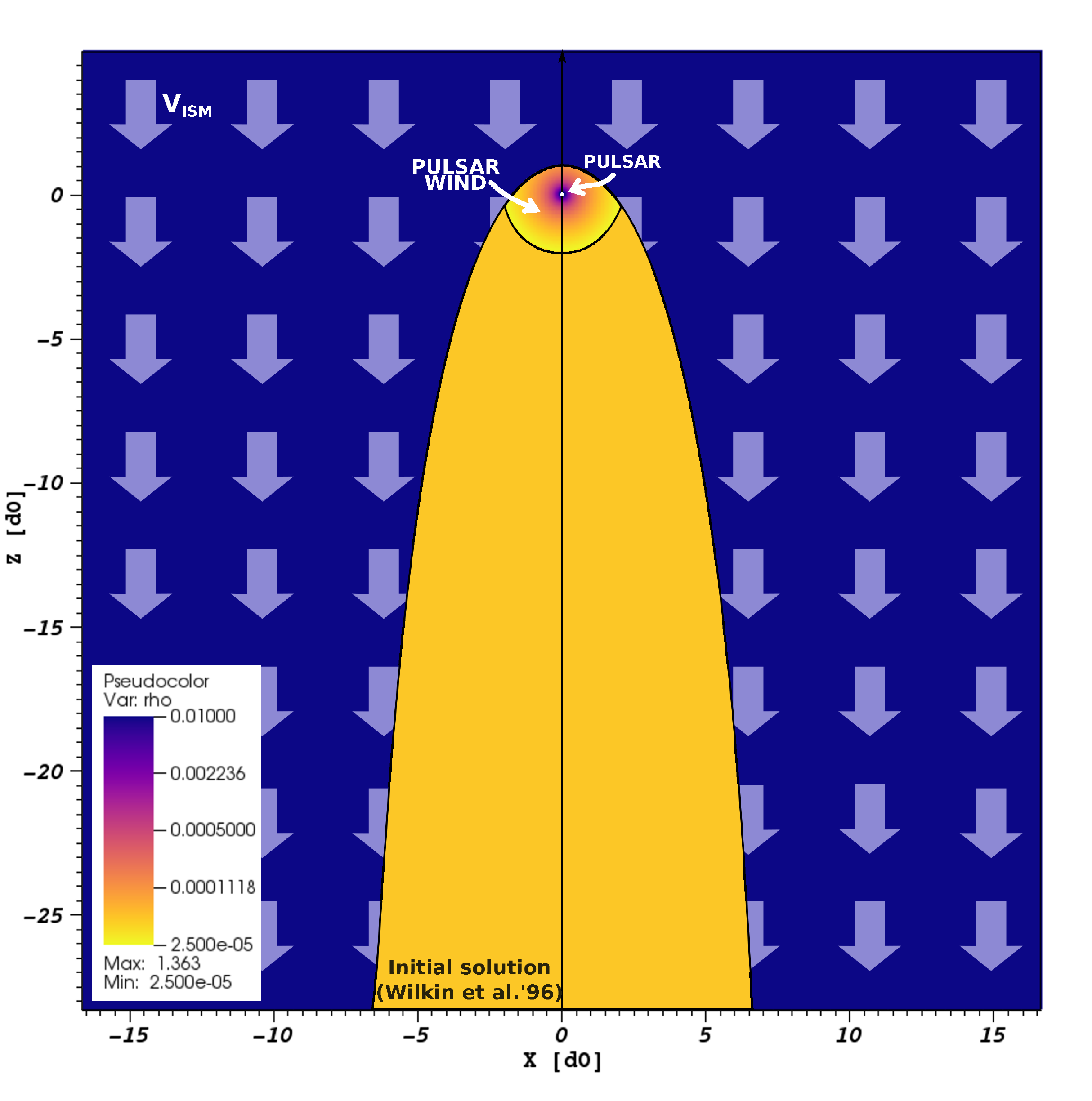}
	\caption{Sketch of the initial setup color-coded with a 2D slice of a density map (in logarithmic scale) of the initial conditions ($t= 0$). The axes are in units of $d_0$. 
	The simulated data are read and elaborated mostly with the use of VisIt \citep{Childs:2012}, an opensource analysis tool. }
	\label{fig:sketch1}
\end{figure}
This scenario was confirmed by numerical simulations in different regimes  \citep{Bucciantini:2002,Bucciantini_Amato+05a,Vigelius:2007,Barkov:2019} and it is illustrated in Fig.~\ref{fig:sketch1}.

BSPWNe have been observed in detail at different wavelengths in the last years, revealing a puzzling variety of structures at the different scales: different morphologies in the bow shock head, different shapes and elongation of the tails, with astonishing outflows misaligned with the pulsar velocity. 
They have been mainly revealed as non-thermal radio and X-ray emission \citep{Arzoumanian_Cordes+04a, Gaensler:2004, Yusef-Zadeh:2005, Li:2005, Gaensler05a, Chatterjee:2005, Kargaltsev_Misanovic+08a, Misanovic_Pavlov+08a, Ng_Camilo+09a, Hales_Gaensler+09a, Ng_Gaensler+10a, De-Luca_Marelli+11a, Ng:2012, Marelli_De-Luca+13a, Jakobsen_Tomsick+14a, Auchettl:2015,  Klingler_Rangelov+16a, Posselt_Pavlov+17a}, with polarimetric informations only available for a few cases, suggesting a large variety of magnetic configurations \citep{Ng:2012, Yusef-Zadeh:2005, Ng_Gaensler+10a, Klingler:2016, Kargaltsev:2017}.
Extended TeV halos have also been recently detected surrounding some BSPWNe \citep{Abeysekara:2017}, attracting much attention because these nebulae are thought to be one of the major contributors of leptonic anti-matter in the Galaxy, in competition with possible dark matter sources \citep{Blasi:2011, Amato:2017}.
If the pulsar is moving through a partially ionized medium the collisional and/or charge exchange excitations of neutral hydrogen atoms in the tail can also allow one to reveal the BSPWN as H$_\alpha$ emission \citep{Chevalier:1980, Kulkarni_Hester88a,Cordes:93,Bell_Bailes+95a,van-Kerkwijk_Kulkarni01a,Jones_Stappers+02a,Brownsberger:2014,Romani_Slane+17a}, or alternatively  in the UV \citep{Rangelov_Pavlov+16a}, and IR \citep{Wang_Kaplan+13a}.

Neutral contamination can also modify the tail dynamics and morphology, as theoretical and numerical models have recently predicted \citep{Morlino:2015, Olmi:2018}. The nature of this changes can be very important as an indicator of the typical speeds in the tail and of the nature of the turbulence that, together with the magnetic configuration, are determinant in understanding how particles escape from these objects \citep{Bucciantini:2018}, fundamental to properly assess the efficiency of pair contamination in the ISM.

The earliest attempts to model BSPWNe date back to more than one decade. Apart from analytical and semi-analytical works \citep{Wilkin:1996, Bandiera93a}, the first numerical models have been presented in the classical hydrodynamical regime by \citet{Bucciantini:2002} and \citet{van-der-Swaluw:2003}.
Due to their intrinsic limitations, first of all the absence of the magnetic field, these models can only account for the properties of the outer layer of shocked ISM, whence the H$_\alpha$ emission comes and where the magnetic field is not expected to be very important, and for the global structure of the system.
The first effort to include magnetic field was done by \citet{Bucciantini:2005}. Here results from a series of 2D MHD relativistic and axisymmetric numerical simulations have been presented, with particular attention to the effects of different  magnetization (the ratio of the magnetic to the kinetic energy fluxes) on the PWN structure, and its emission signatures.
The authors found that the level of magnetization does not influence the external layer of the shocked ISM, confirming the previous findings of HD models. The growth of magnetization causes an enhanced emission near to the symmetry axis and modulations of emissivity are also observed in association with shear instabilities in the nebula.

The axisymmetric assumption of those simulations however forced the authors to consider only the case of a pulsar spin axis aligned with the pulsar kick velocity (and also to include ad hoc magnetic dissipation in the bow shock head to avoid numerical artifacts related to the enforced symmetry of the system). 
However there is no reason to expect such alignment \citep{Johnston_Hobbs+05a,Ng_Romani07a,Johnston_Kramer+07a,Noutsos_Kramer+12a,Noutsos_Schnitzeler+13a}. 
To account for the spin-kick inclination one needs to work in full 3D.
The third dimension is particularly important for the correct description of the magnetic field configuration \citep{Bucciantini:2017}, and it is crucial in the study of the development of turbulence, which can be strongly affected by geometric constraints. 
Limits of 2D models of PWNe in that sense have been largely discussed in the last years \citep{Del-Zanna:2006, Olmi:2013, Olmi:2014, Porth:2013}, since the first 3D MHD models of PWNe become available \citep{Porth:2014, Olmi:2016}. 
They confirm that the full 3D modeling allows the development of a complex structure of the magnetic field which is artificially prevented in 2D, due to geometrical limitations.
Moreover, the growth of 3D instabilities leads to an higher magnetic dissipation, such that higher values of the injected magnetization can be reached \citep{Porth:2013}. 
2D models were proved to be still robust in reproducing the properties of the inner region of PWNe \citep{Del-Zanna:2006,Volpi:2008,Camus:2009,Olmi:2014}, since deviations from these simplified models become mainly important in the outer region of the PWN. 
This might however not be satisfactory  in the case of BSPWNe, where the tail regions extend far away from the pulsar location.

The first attempt to 3D modeling of BSPWNe has been done by \citet{Vigelius:2007} but limited to the classical HD regime. They present a series of simulations of bow shocks not very extended in the tail region, considering different latitude variations in the pulsar wind and various density gradients in the ISM. In particular they found that the wind anisotropy does not directly influence the bow shock morphology in the head, which is on the contrary more deeply affected by the interaction with the anisotropic ISM. 
More recently 3D MHD simulations have been presented in \citet{Barkov:2019}. 
Different inclinations of the magnetic field with respect to the pulsar spin axis, and various magnetizations are considered, together with a possibly non uniform  external density of the ISM. The authors conclude that the magnetic field has a fundamental influence in accounting for the bow shock morphology, while anisotropies in the ISM density only lead to marginal variations of the external surface \citep{Barkov:2019, Toropina:2018}. 
However some of the choices in these models (the initial wind Lorentz factor is less than 3,  TS is not fully detached from the simulation boundary) raise a few questions of the robustness of the results.

Aim of the present work is to investigate in the full 3D relativistic MHD regime the morphology of BSPWNe, varying the pulsar parameters: inclination of the magnetic axis with respect to the pulsar spin axis, anisotropy of the wind momentum flux and level of magnetization. We are in particular interested in determining how the large scale structure of the bow shock and its tail are influenced by the morphology of the magnetic field and by the level of turbulence. 
In the present paper we will then focus our attention on the analysis of the dynamics. 
A detailed analysis of the emission, polarimetry and variability properties would be postponed to successive papers.

%
This paper is organized as follows: in Sec.~\ref{sec:Nsetup} the physical model of the pulsar wind, the numerical tool and setup used for our simulations are described; in Sec.~\ref{sec:results} we present and discuss our findings. Conclusions are finally drawn in Sec.~\ref{sec:conclusion}.
\section{Pulsar wind model and numerical setup}
\label{sec:Nsetup}
Our simulations have been performed with the numerical code PLUTO \citep{Mignone:2007}. PLUTO is a shock-capturing, finite-volume code for the solution of system of hyperbolic and parabolic partial differential equations, particularly oriented to astrophysical fluid dynamics. 
It supports the use of adaptive mesh refinement (AMR) thanks to the CHOMBO libraries, a fundamental requirement for the present study \citep{Mignone:2013}. The numerical grid must in fact account for a sufficient resolution to capture simultaneously the pulsar wind injection region and the large scale of the cometary tail of the pulsar bow shock nebula.

A second order Runge-Kutta time integrator and an HLLD Riemann solver (the Harten-Lax-van Leer for discontinuities, see \citealt{Miyoshi:2005, Mignone:2006}) have been used, in order to better treat shear layers in the nebula. 
The HLLD solver is automatically relaxed to the HLL one in the regions of the domain occupied by the relativistic pulsar wind, in order to be able to use a Lorentz factor $\gamma=10$, high enough to ensure that the TS jump conditions are in the relativistic regime, even in those region of the TS that are strongly inclined with respect to the wind itself. This has been made possible due to a customization of the code.
The solenoid condition of the magnetic field, $\mathbf{\nabla}\cdot \mathbf{B}=0$ is maintained using a divergence-cleaning method \citep{Dedner:2002}.

The typical length scale of BSPWNe is the so called {\it stand off distance} $d_{0}$  \citep{Wilkin:1996,Bucciantini:2001, van-der-Swaluw:2003, Bucciantini_Amato+05a}, where the wind momentum flux and the ISM ram pressure balance each other
\begin{equation}\label{eq:stagnationp}
	d_{0} = \sqrt{\dot{E}/(4\upi c \rho_\mathrm{ISM} v^2_\mathrm{PSR})}\,,
\end{equation}
where $\dot{E}$ is the pulsar spin down luminosity, $\rho_\mathrm{ISM}$ is the ISM density, $ v_\mathrm{PSR}$ the speed of the pulsar with respect to the local medium, and $c$ the speed of light.  
The typical values of the stand-off distance are of the order of $\sim 10^{16}-10^{17}$ cm \citep{Chevalier:1980, Kulkarni_Hester88a, Cordes:93, Chatterjee:2002, Romani:2010, van-Kerkwijk_Kulkarni01a, Brownsberger:2014,Romani_Slane+17a}. Similarly a convenient choice for parametrizing time in the inner tail is the characteristic flow time $t_0 = d_{0}/c$, while the relaxation time with respect to the interaction with the ISM is typically $t = d_{0}/ v_\mathrm{PSR}$.

\begin{figure}
	\centering
	\includegraphics[width=.5\textwidth]{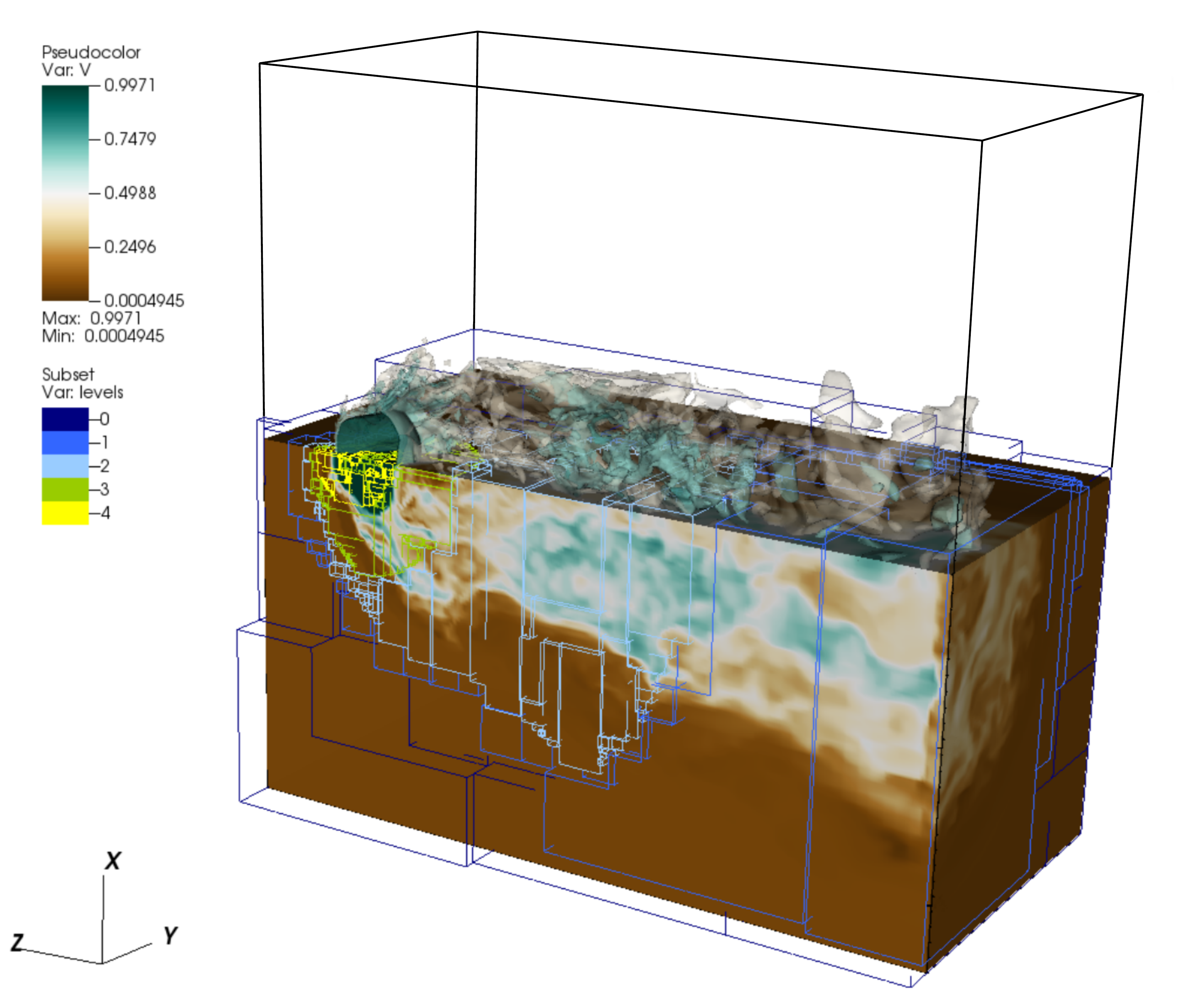}
	\caption{ Velocity magnitude in a composite image for run A$_{\{ 0,1\}}$. In the lower part of the picture the velocity is shown as standard intensity plot with 2D cuts on orthogonal planes. In the upper part the same variable is represented with transparent isocontours at ten uniform spaced levels along the same color scale. Contours of the AMR levels are over imposed with different colors representing the boundaries of AMR boxes.
	}
	\label{fig:B45lev}
\end{figure}

The simulation domain in cartesian coordinates is centered on the pulsar with the $z$-axis aligned with the pulsar kick velocity (the reference frame is moving with the pulsar, thus the ISM is seen as a uniform, unmagnetized flow moving along the negative $z$ direction). 
The domain extends in the range $[-17 d_0,\, +17 d_0]$ along the $x$ and $y$ direction and $[-28 d_0,\, 5 d_0]$ along $z$.
This choice have been made in order to allow for the development of a long tail behind the pulsar and to reduce as much as possible the domain occupied by the un-shocked ISM. The base grid have $128^3$ equally spaced grid points. A set of four AMR levels is then used in order to reach the required resolution around the pulsar, in order to resolve at best the bow shock head, which is a strongly dynamic region, corresponding to an effective resolution of $2048^3$ cells at the highest level.
The configuration of AMR levels can be seen in Fig.~\ref{fig:B45lev}.

In order to speed up the simulation towards a steady relaxed regime, at the beginning the domain is divided into two regions with separatrix given by the analytical bow shock shape \citep{Wilkin:1996}
\begin{equation}\label{eq:wilkin}
	R_w(\theta) = d_0 \csc \theta \sqrt{3(1-\theta \cot\theta)}\,,
\end{equation}
where  $(r,\,\theta,\, \phi)$ are the spherical coordinates, centered on the pulsar, and with the polar axis aligned to the $z$ direction. In $r > R_w$ the domain is initialized with a cold slowly moving ISM, with a speed  $v=-v_{PSR}=0.1 c$. 
In $r \le R_w$ and $r > 2d_0$ we set a uniform density and pressure moving backward with a speed  $v=-v_{PSR}=0.1 c$ to allow the bow shock tail to develop without the eventual formation of back-flows towards the pulsar. 
The value of the PSR velocity in the ISM has been set in order to speed up the relaxation, thus imposed to be $v_{PSR}=0.1 c$, in the $-z$ direction.
This value was shown to be small enough to recover the proper dynamics of the non-relativistic ambient medium \citep{Bucciantini:2002, Bucciantini:2005}.
In $r \le R_w$ and $r \le 2d_o$ we initialize the relativistic pulsar wind, with radial velocity $v_r=(1-1/\gamma_0^2)^{1/2}/c$. 
For simplicity, given that we are here mostly interested in the dynamics of the relativistic component, the ISM is assumed to be unmagnetized. The external magnetic field will be important for the escape of particles from the bow shock into the ambient medium, while being energetically subdominant in the ISM, its effect on morphology of the bow shock nebula will be negligible.
This setup is shown in  Fig.~\ref{fig:sketch1}, where a 2D slice of the density profile illustrates the initial conditions. 
The pulsar wind is continuously injected from a radius $r_\mathrm{inj}\simeq0.2 d_0$. Such value for the injection boundary ensures that the TS radius is always detached during the entire evolution.
The possibly anisotropic injection of energy from the pulsar wind is accounted for with a modulation of the wind density
\begin{equation}\label{eq:in_rho}
	\rho =
   		\begin{cases}
   		(\rho_0/r^2)  \mathcal{F}(\psi)  			&  \quad \mathrm{if} \quad r\le 2 d_0 \,{\rm and}\; r\le 2 R_w(\theta) \, ;\\
   		\rho_0/d_0^2	 				 	& \quad \mathrm{if} \quad  r\ge 2 d_0  \,{\rm and}\; r\le R_w(\theta) \, ;\\
   		\rho_\mathrm{ISM}					& \quad \mathrm{otherwise}\,,
		\end{cases}
\end{equation}
with $\rho_0=[v_\mathrm{PSR} d_0 /(c \gamma_0)]^2 \rho_\mathrm{ism}$.
The wind Lorentz factor is kept constant at the value $\gamma_0=10$,  high  enough to ensure that the post-shock dynamics is independent on its value \citep{Del-Zanna:2004}.
Modulation is governed by the function $\mathcal{F}(\psi)$, connected to the wind energy flux by the expression
\begin{equation}\label{eq:flux}
	\mathcal{F}(\psi) = N_0\left(  1 + \alpha \sin^2 \psi \right)\,,
\end{equation}
where $\psi$ is the colatitude of the generic point $(x_0,\,y_0,\,z_0)$ from the pulsar spin axis, defined from the relation $\cos{\psi}=(y_0 \sin{\phi_M} + z_0 \cos{\phi_M})/r_0$, where $r_0^2=(x_0^2+y_0^2+z_0^2)$ and $\phi_M$ is is the spin-kick inclination (configurations with $\phi_M=0$ have the spin axis aligned with the bow shock symmetry axis $z$). $N_0$ is a normalization factor, fixed by the requirement $\int_\Omega \mathcal{F}(\psi) d\Omega =4\upi$ and $\alpha$ a dimensionless parameter governing the level of anisotropy of the wind, with the isotropic distribution recovered for $\alpha=0$.

The pressure in the wind is simply given by entropy conservation with adiabatic index $\Gamma$, while is set fixed the other regions
\begin{equation}\label{eq:in_prs}
	p  =
   		\begin{cases}
   		  p_0\left(\rho/\rho_0 \right)^\Gamma 		& \quad \mathrm{if} \quad r\le 2 d_0 \,{\rm and}\; r\le 2 R_w(\theta) \, ;\\
   		 \rho_0/[ 6 ( d_0 \gamma_0)^2]  			& \quad \mathrm{if} \quad  r\ge 2 d_0  \,{\rm and}\; r\le R_w(\theta) \, ;\\
   		 p_\mathrm{ISM}					& \quad \mathrm{otherwise}\,,		 
		\end{cases}
 \end{equation}		
with $p_0=0.01\rho_0 (c/d_0)^2$ and $\Gamma=4/3$, the appropriate value for the relativistic shocked pulsar plasma. 
In \citet{Bucciantini:2002} more sophisticated simulations with a multi-fluid treatment were realized: different equations of state are considered for describing the relativistic component of the wind and the non-relativistic material of the ISM. The authors found that finally this only leads to minor deviations to the overall geometry of the bow shock. 
A comprehensive treatment of the two components can be also achieved by the use of adaptive equations of state, like Taub's one \citep{Taub:1948, Mignone:2005}. However they should not be used in the presence of strong mass contamination, since they are based on the assumption of instantaneous thermalization between the two components. This can easily happen in turbulent flow due to numerical diffusion, but physically this is expected to happen on timescales much longer than the typical flow time in the tail of a BSPWN, given the mass and energy difference of the two components. 

In the wind the magnetic field in a generic point $(x_0,y_0,z_0)$, 
is defined in cartesian coordinates as 
\begin{equation}\label{eq:in_B}
	\mathbf{B}  = \left(\frac{B_0 \sin \psi}{ \mathcal{R}}\right)\left(\frac{d_0}{r_0} \right)\times 
		\begin{cases}
   		   		 \left(-y_0 \cos{\phi_M} + z_0 \sin{\phi_M} \right)   \mathbf{e_x}\,;\\
				 \left(x_0\cos{\phi_M}\right)   \mathbf{e_y}\,;\\
				 -\left(x_0 \sin{\phi_M}\right)    \mathbf{e_z}\,;\,
		\end{cases}
 \end{equation}		
where $\mathcal{R}=(x_0^2 + y_0^2\cos^2{\phi_M} + z_0^2\sin^2{\phi_M} -y_0z_0\sin{2\phi_M})^{1/2}$. 
The previous formula is applied in the region $r\le 2 r_b$, while the magnetic field is imposed to be 0 in the rest of the domain.

 The runs are summarized in table \ref{tab:tabRuns}. We investigated various magnetizations, with $\sigma=[0.01,1, 0.1,\, 1.0]$, and inclination of the pulsar spin axis $\phi_M=[{0,\,\upi/4,\, \upi/2}]$, both for isotropic and anisotropic distribution of energy flux in the wind. 
 In order to speed up simulations we evolve a pure hydrodynamic isotropic case up to $t_{H,f}=740 t_0$, sufficient for the configuration to be completely relaxed. This is then used as the starting point for all the other runs. 
Anisotropy and wind magnetization are gradually increased to avoid spurious jumps in the injection conditions. 
Once the injection conditions have reached their desired final value, the system is evolved again up to a quasi-stationary configuration until the time $t_f=1152 t_0$, enough for the dense slow ISM material to stream from the head of the bow shock to the far tail. We have verified in each case that at $t_f$ the tail is relaxed to a final state.
\begin{table}
\begin{center}
\begin{tabular}{{ccccccccc}}
\hline
 Run  			& Dimensions	&	$\alpha$ 		 &	 $\phi_M$ 		& $\sigma^\mathrm{ [a]}$      & $t_f \,[t_0]$   \\
\hline
 2D-H$_0$		& 2D			&	0 			& 	0  			& 0        		&	1000	 \\
 H$_0$			& 3D			&	0 			& 	0  			& 0      		&	740	 \\
 I$_{\{0,i\}}$		& 3D			&	0 			& 	0 			& $10^{-i}$       &	1152	 \\
 I$_{\{\upi/4,i\}}$	& 3D			& 	0			& 	$\upi/4$  		& $10^{-i}$        &	1152	 \\
 I$_{\{\upi/2,i\}}$	& 3D			& 	10 			& 	$\upi/2$  		&  $10^{-i}$       &	1152  \\
 A$_{\{0,i\}}$		& 3D			&	10 			& 	0  			&  $10^{-i}$       &	1152	 \\
 A$_{\{\upi/4,i\}}$	& 3D			&	10			& 	$\upi/4$ 		&  $10^{-i}$       &	1152	 \\
 A$_{\{\upi/2,i\}}$	& 3D 		&	10			& 	$\upi/2$  		& $10^{-i}$       &	1152	 \\
\hline
\multicolumn{6}{l}{%
  \begin{minipage}{6.5cm}%
    \tiny{ [a] For MHD runs values of the magnetization are given by  the exponent $i=[2,1,0]$, such that $\sigma=[10^{-2},\, 10^{-1},\, 10^{0}]$}.%
  \end{minipage}%
}\\
\end{tabular}
\end{center}
\caption{List of the runs. Subscript $i$ in the run name indicates the value of the initial magnetization (in the exponent notation specified in the Table note).}
\label{tab:tabRuns}
\end{table}

\section{Results and Discussion}
\label{sec:results} 
In this section we analyze the dynamics resulting from the different configurations listed in Table \ref{tab:tabRuns}.
Notice that all the quantities are always given in code units, that can be rescaled to physical units in terms of $d_{0}$, defined by eq.~\ref{eq:stagnationp}. 
As a useful quantity for converting between the two we chose the external ram pressure, which in code units for us is $p_\mathrm{ext} = \rho_\mathrm{ISM}v_\mathrm{PSR}^2=0.01$. 
The typical ram pressure for bow shocks in cgs units is $p_\mathrm{ext} = 10^{-10} \rho_{24} v_7^2$ g cm$^{-1}$ s$^{-2}$, where density is expressed in units of $1\times10^{-24}$ g cm$^{-3}$ and velocity in units of $10^{7}$ cm s$^{-1}$.
 
 We begin by comparing the large scale structure of bow shock in 2D versus 3D. 
\begin{figure*}
	\centering
	\includegraphics[width=.99\textwidth]{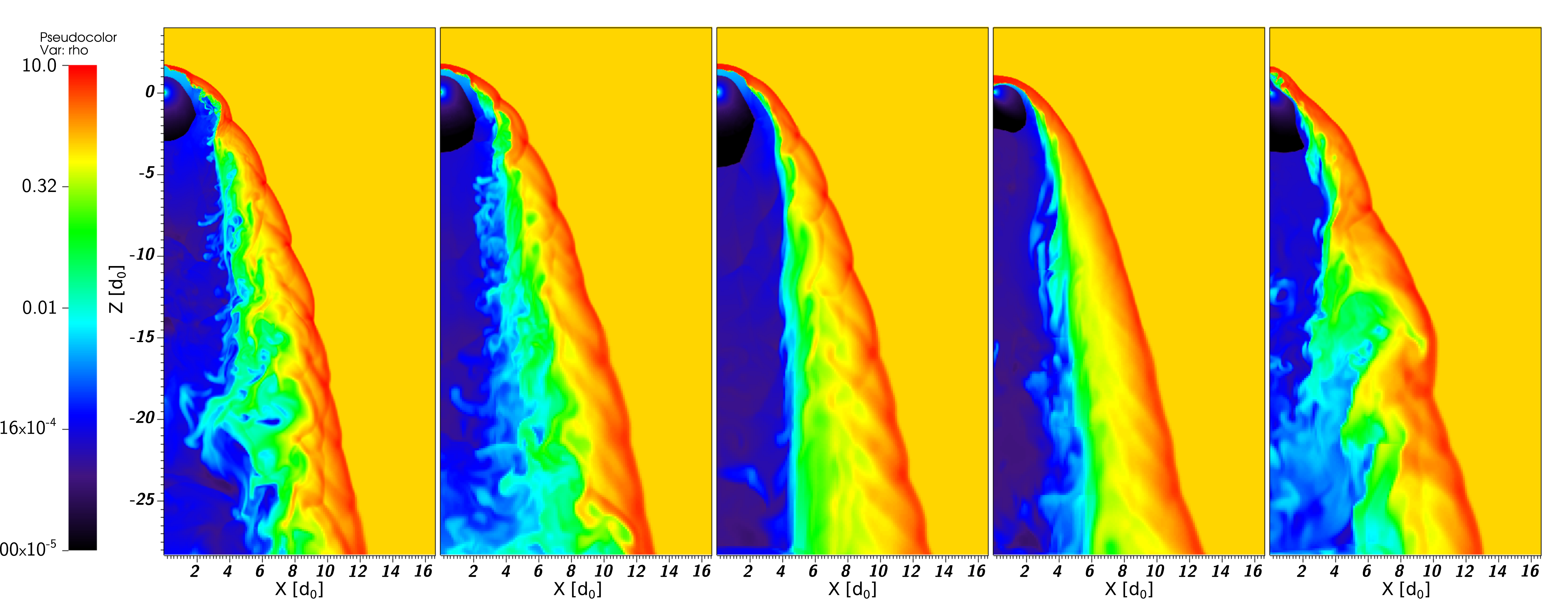}	
	\caption{Planar slices of density maps, in logarithmic scale and code units. Maps from different runs are compared at an equivalent stage of evolution of the system, where the configuration is completely relaxed. From left to right maps refer to runs: 2D-H$_0$, H$_0$, I$_{\{ 0,0\}}$, A$_{\{ 0,0\}}$ and A$_{\{ \upi/4,0\}}$.}
	\label{fig:Densities}
\end{figure*}
%
An overview image comparing different runs from 2D to 3D and from HD to MHD can be found in 
Fig.~\ref{fig:Densities}, where density profiles are shown as 2D slices in the $(x,\,z)$ plane for runs, from the left to the right-side, 2D-H$_0$, H$_0$, I$_{\{ 0,\,0\}}$, A$_{\{0,\,0 \}}$  and A$_{\{ \upi/4,\,0\}}$. Here we can notice a few major differences from case to case: HD runs appear to be more affected by small scales turbulence, while MHD cases (all with $\sigma=1.0$), show a global structure much less influenced by turbulence. As we will see in details in the following, this is a characteristic of high-magnetized cases, in which the properties of the fluid are usually dominated by conditions at injection.
The other clear difference is the shape and dimension of the TS, which as expected is very different from the isotropic to the anisotropic model, where it assumes an oblate shape.
The outer shape of the bow shock instead appears to be quite similar from case to case. The only exception is the right-most plot, referring the case with $\upi/4$ spin-kick inclination, which shows an evident variation of the FS shape. As we will see in the following it is also the only one characterized by an evident asymmetry around the $z-$axis.
 
\begin{figure}
	\centering
	\includegraphics[width=.495\textwidth]{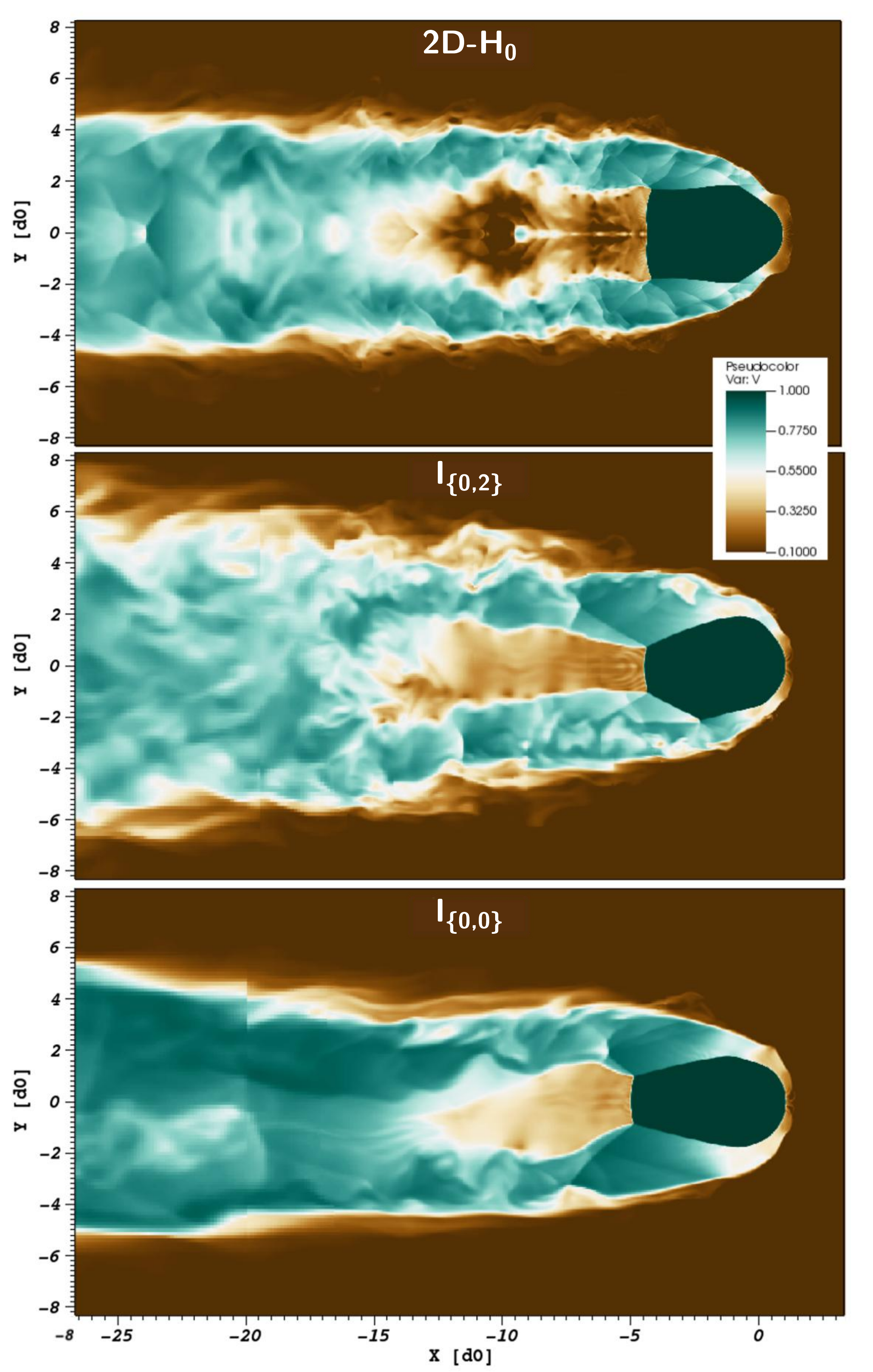}
	\caption{Comparison of the 2D maps of the velocity magnitude for runs 2D-H$_0$, I$_{\{0,2 \}}$ and I$_{\{0,0 \}}$ at $t=t_f$. Velocity is given in units of $c$. }
	\label{fig:cfr_V}
\end{figure}
In Fig.~\ref{fig:cfr_V}, the velocity magnitude in the 2D-H$_0$ case, at $t=t_f$, is directly compared with 2D slices from runs I$_{\{0,2 \}}$ and I$_{\{0,0 \}}$ in the $x=0$ plane. 
These 3D runs have a spin axis aligned with the kick axis, and thus the injection conditions are fully axisymmetric. 
The shape and size of the TS are essentially the same: the location of the TS in the front region, as well as the distance of the Mach disk on the backward side, show no significative difference. 
3D simulations show however a TS having a drop-like shape more than a bullet-shape as in 2D, partly due to the collimating effect of the magnetic field (a toroidal field wounded around the $z-$axis), and partly due to the development of turbulence in the tail. 
It is interesting to observe that, while in 2D there are several inclined weak shocks in the tail, those tend to disappear in 3D,  where a more turbulent flow develops.
The behavior of turbulence however is strongly dependent on the magnetization of the wind. 
High values of $\sigma$ lead to flow structures in the tail that are more laminar, and this has two consequences. 
On one hand the turbulent ram pressure lead to a sideway expansion of the tail, which  is more pronounced in the I$_{\{0,2 \}}$ run, while the  I$_{\{0,0 \}}$ has a tail cross section comparable to the 2D HD result. On the other the development of turbulence slows down the bulk flow in the tail. 
Typical values of the flow speed in the tail ranges from $0.6c$ in the I$_{\{0,2 \}}$ to $0.9c$ in I$_{\{0,0 \}}$ run. 
One also sees that in all cases, the inner slow channel that was found in early 2D simulations \citep{Bucciantini:2004} does not extend more than $5 d_0$ in the tail, and is completely lost at distance larger than  $10 d_0$.

\begin{figure}
	\centering
	\includegraphics[width=.45\textwidth]{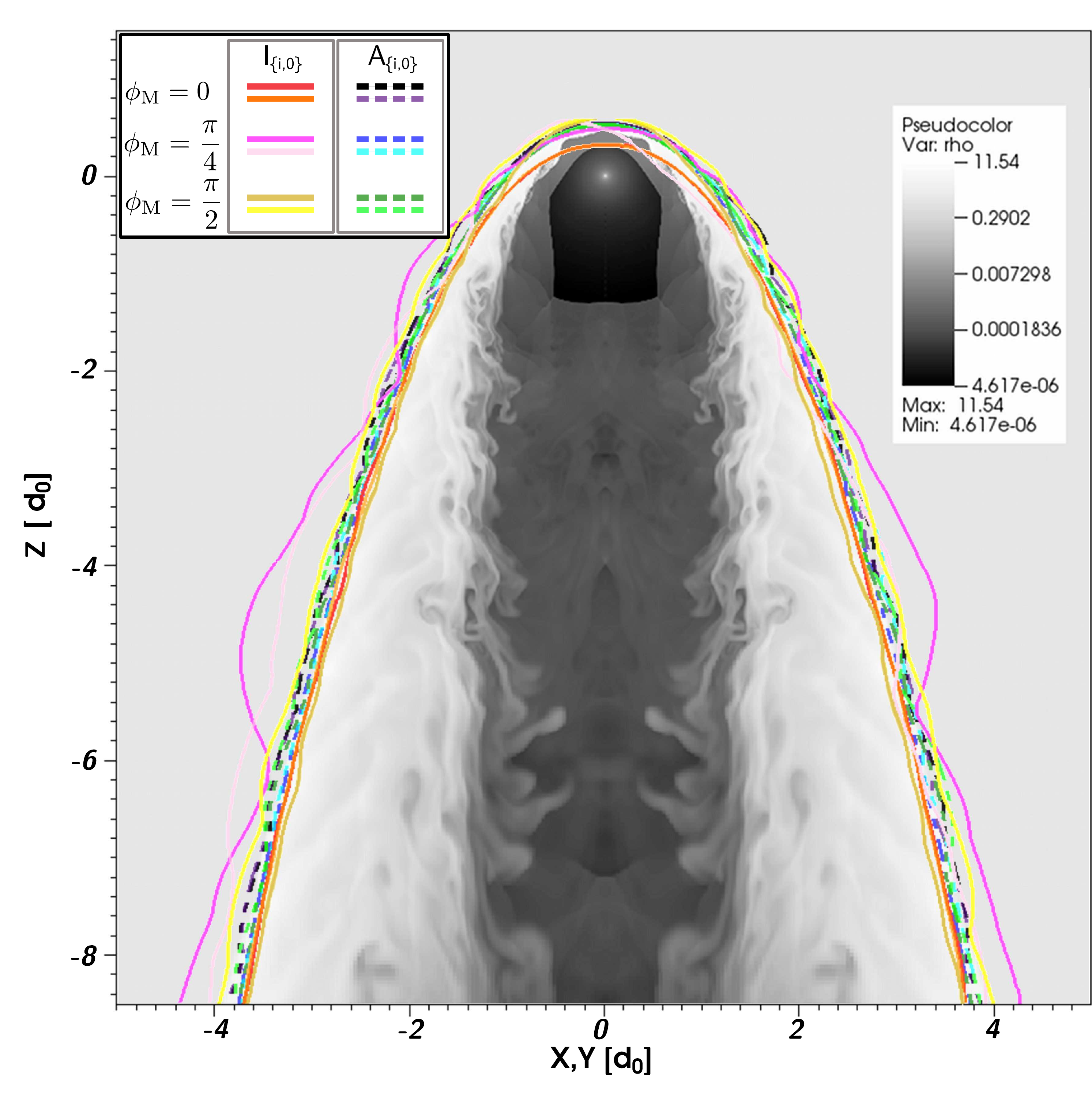}
	\caption{Contour lines of the forward shock referring to the case of $\sigma=1.0$ for all the considered configurations. 
	Contours are superimposed on a density map from run 2D-H$_0$, to show the similarity of the FS from 2D to 3D simulations.
	Solid lines (warm colors) and dashed lines (cool colors) indicate the isotropic and anisotropic wind cases respectively. Different inclinations of the magnetic field (i.e. values of $\phi_M$) are represented with different colors, as specified in the plot legend. 
	Contours are obtained from 3D datasets at $t=t_f$ by extracting values from two orthogonal slices (for $x=0$, the upper colored line of each set, and $y=0$, the bottom line) for each case.  
	}
	\label{fig:FS}
\end{figure}
We have also compared the shape of the forward bow shock among all our runs. 
This can be seen in Fig.~\ref{fig:FS}, where the position of the FS is given as a contour line of the density for all the configurations with $\sigma=1.0$. Contours are superimposed on a density map from the 2D hydrodynamic case, for direct comparison. 
Here the forward shock shape and position appear to be very similar from case to case, with the major deviations coming from case A$_{\{\upi/4,0\}}$, mostly because the inclination leads to the development of waves that propagate backward modulating the shock shape.
 A sort of lateral blobs are visible at the FS surface, possibly representing periodic structures. 
 In any case their extension is in general smaller than $d_0/2$, or $\sim 1d_0$ at most, and should probably be completely invisible to actual instruments, due to resolutions limits.
This confirms that the outer shape of a BSPWN appears then to be quite independent on the properties of the pulsar wind, even at high magnetization, but rather dominated by the interaction with the ambient medium, as predicted by previous models \citep{Romani:1997, Vigelius:2007}. 

We also found that the distance between the bow shock and the contact discontinuity (which is usually named $\Delta$) is in perfect agreement with the predicted value of $5/16d_0$  \citep{Chen:1996} for the isotropic model, while in the anisotropic case it is more variable due to the different morphology of the shock, and it goes from 5/16 up to 5/8, depending on the magnetic inclination.

\begin{figure}
	\centering
	\includegraphics[width=.5\textwidth]{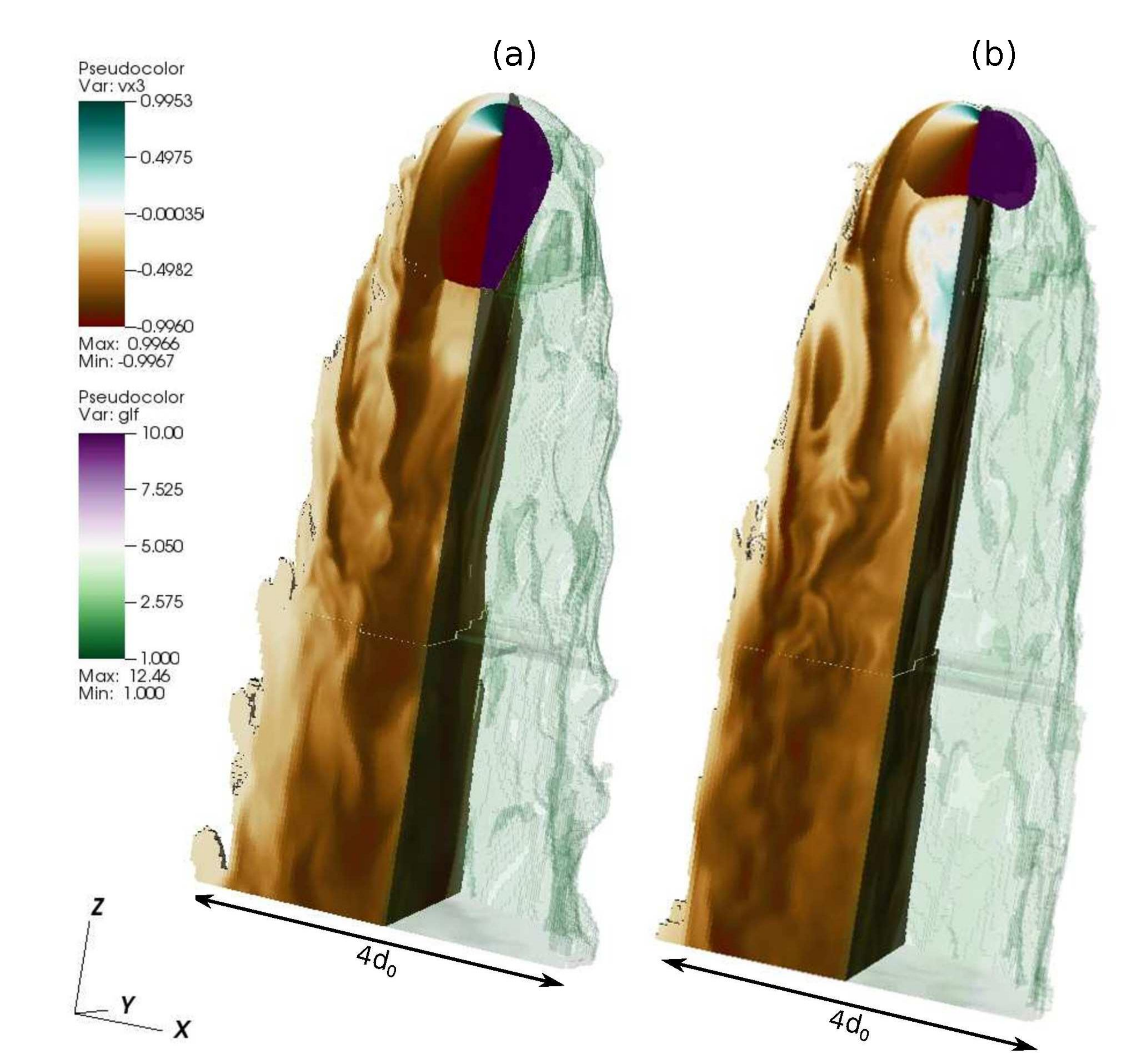}\\
	\caption{3D combined color maps of the $v_z$ profile (left-half  of each map) and wind Lorentz factor $\gamma$ (right-half of each map). 
	Letter (a) identifies run I$_{0,0}$ and letter (b) run A$_{0,0}$. }
	\label{fig:cfrVz_GLF}
\end{figure}
The effect of an anisotropic energy distribution in the pulsar wind is shown in Fig.~\ref{fig:cfrVz_GLF} where the left-half side of each map shows the $v_z$ component of the velocity in a cut of the 3D dataset, while in the right-half side of the map the Lorentz factor is drawn, with the purple area indicating the shock surface (at value $\gamma=10$). 
As expected in the anisotropic model, where the energy flux in the wind is not uniform, the TS assumes an oblate shape, with major extension in the direction in which the energy flux peaks (i.e. the direction orthogonal to the magnetic axis). In this sense the shock tends to reflect more the structure of the energy injection than the motion of the PSR. Here the elongation of the TS into a bullet shape is lost.
The deformation of the shock is largely responsible for the formation of turbulence in the anisotropic runs: the bow shock head appears to be more dynamic, with vortexes on smaller scales and less coherent structures. 
The region in which the high velocity channel streams along the oblate part of the shock appears to be thinner, and randomization of the flow causes this structures to dissolve near to the head region. 
Moreover the different shape of the shock also causes an important change in the zone occupied by the slow inner channel, which now turns into a ``stagnation zone'' of the flow behind the TS. 
\begin{figure}
	\centering
	\includegraphics[width=.55\textwidth]{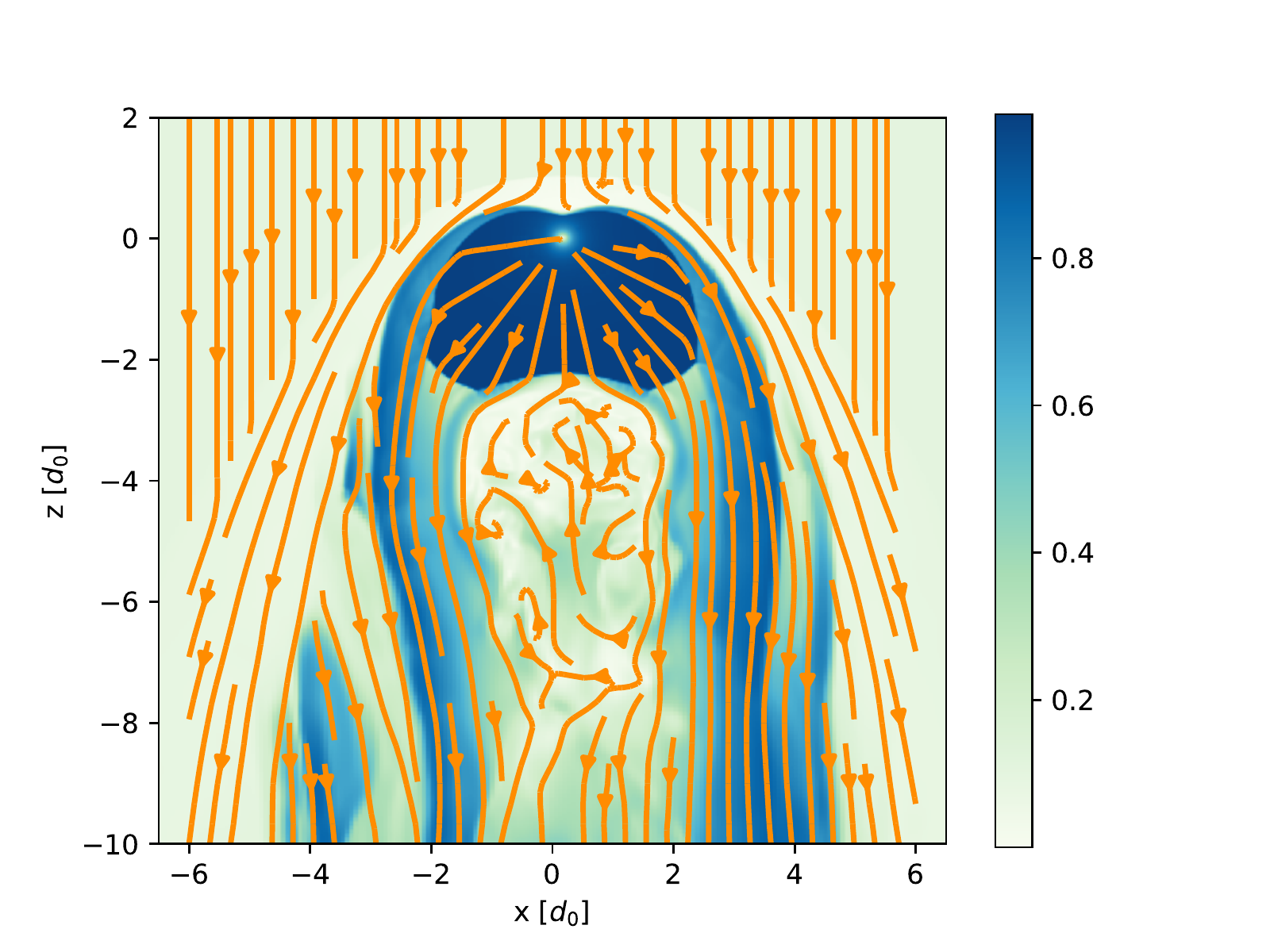}\\
	\caption{Map of the velocity magnitude with streamlines from a 2D slice in the $x-z$ plane of run A$_{0,0}$. Arrows indicate the direction of the flow and the formation of the ``stagnation zone''  discussed in the text appears more easily recognizable. The color map is expressed in units of $c$.}
	\label{fig:streamL}
\end{figure}
The formation of this ``stagnation zone'' is caused by the material that is first diverted backward to the shock front in the $+z$ direction, then, due to the mixing of the flow in the region behind the shock, pushed on the back of the shock and  diverted again back to the tail, in the $-z$ direction. 
This can be appreciated in Fig.~\ref{fig:streamL} as the effect of the material flowing along the shock front and then squeezed towards the axis around $z=-6$, giving rise to the formation of a turbulent region behind the shock. The map shows a 2D slice of the velocity magnitude and streamlines from run A$_{0,0}$, which shows a much extended ``stagnation zone'' than the isotropic case, visible as a light-blue bubble behind the shock in Fig.~\ref{fig:cfrVz_GLF}.
This region is more extended in the anisotropic case due to the different shape of the shock front, which causes more mixing of the flow behind the shock and in the tail. This also reflects in a more turbulent motion in the tail, causing the velocity to lower and the tail to broaden. 
This effect is not seen in 2D simulations, more likely because to the enforced axisymmetry of those models, which prevent the accumulation of material along the axes.

\begin{figure}
	\centering
	\includegraphics[width=.495\textwidth]{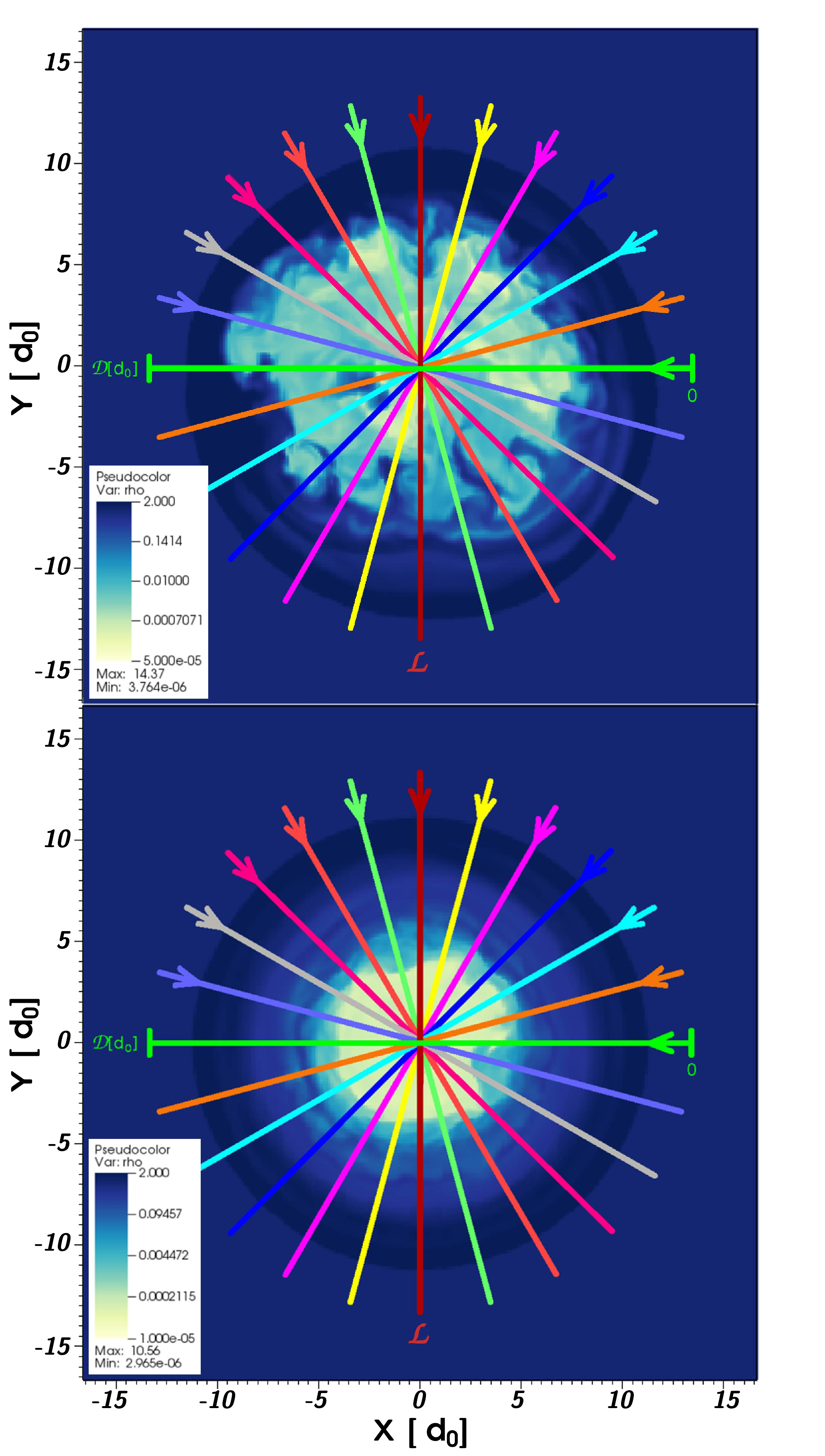}\\
	\caption{Representation of the geometry used for the extraction of 1D profiles of the dynamic variables. 
	We show here the logarithmic maps of the density for run A$_{\{\upi/4,0\}}$ (top) and  run I$_{\{0,0\}}$ (bottom), both at $t=t_f$ and in a 2D slice corresponding to $z=-20.0 d_0$. 
	Colored lines represent the diameters used for extraction of the profiles from each 2D slice, with arrows indicating the direction along the diameter $\mathcal{D}$, ranging from 0 to $\sim 27d_0$. Letter $\mathcal{L}$ indicates the plane that contains the pulsar spin axis. }
	\label{fig:LinOuts}
\end{figure}

\begin{figure*}
	\centering
	\begin{subfigure}{}
	\includegraphics[width=1\textwidth]{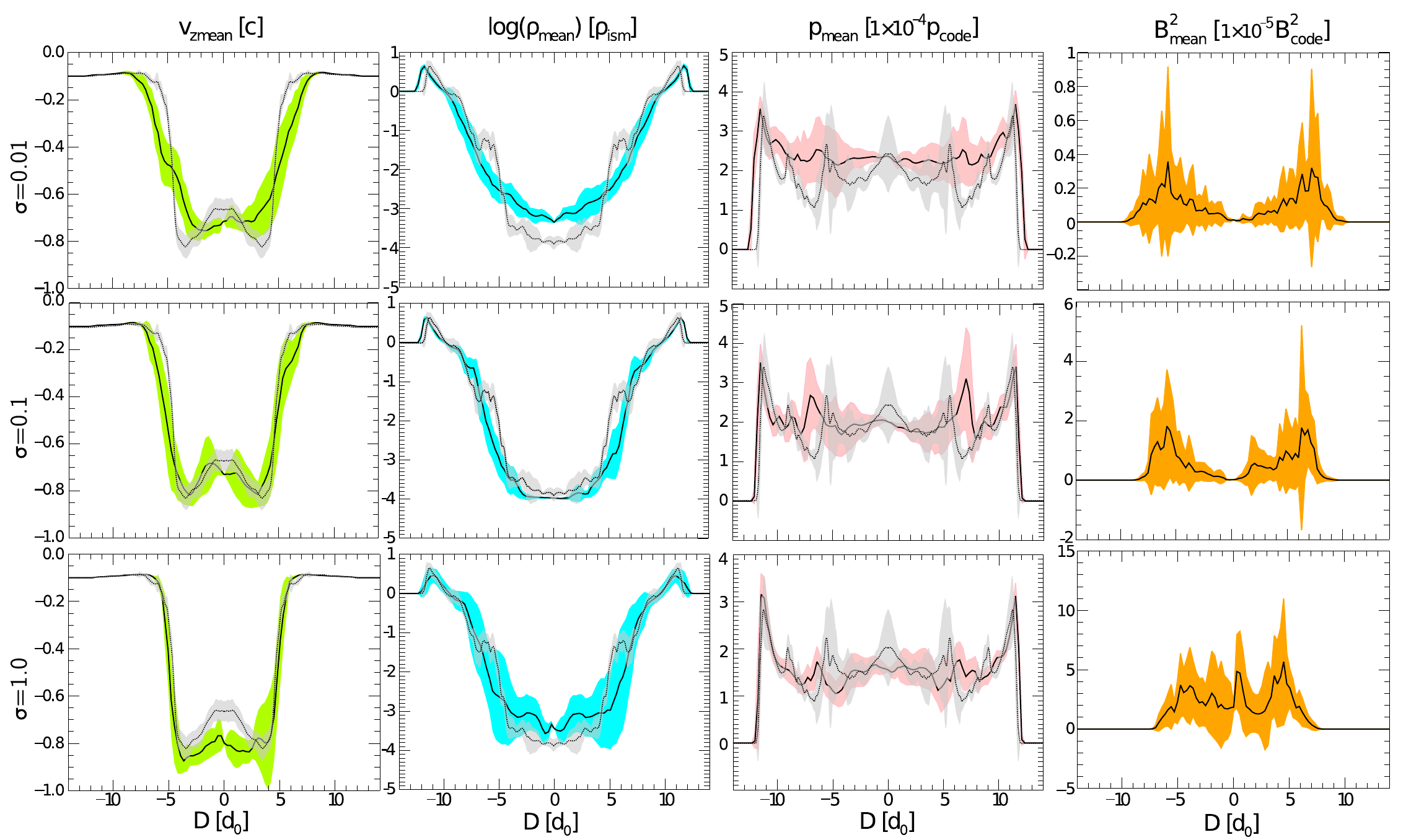}
   	 \caption{Comparison of the average 1D profiles extracted from different 2D maps at various distances from the pulsar (namely 
	 $z=-20.0d_0,\, z=-20.8d_0,\, z=-21.7d_0$) for run I$_{\{0,i\}}$. Columns refers to the same variable (from left to right: $v_z,\, \log{\rho},\,p,\,B^2$) and rows to the same magnetization (from top to bottom $\sigma=0.01,\, \sigma=0.1,\, \sigma=1.0 $).
	Black solid lines indicate the computed average trend of each variable and the surrounding colored area the related standard deviation. The superimposed black dotted line and grey areas are are profiles and standard deviations extracted from 2D-H$_0$, for comparison. 
	Spatial scales are given in $d_0$ units. Notice that $p$ is given in $1\times10^{-4}p_\mathrm{code}$ units and $B^2$ in $1\times10^{-5}B^2_\mathrm{code}$.
	}
	\label{fig:cfr_B0_ISO}
   	 \end{subfigure}\hfill
	\vspace{-0.6cm}
	\begin{subfigure}{}
	\includegraphics[width=1\textwidth]{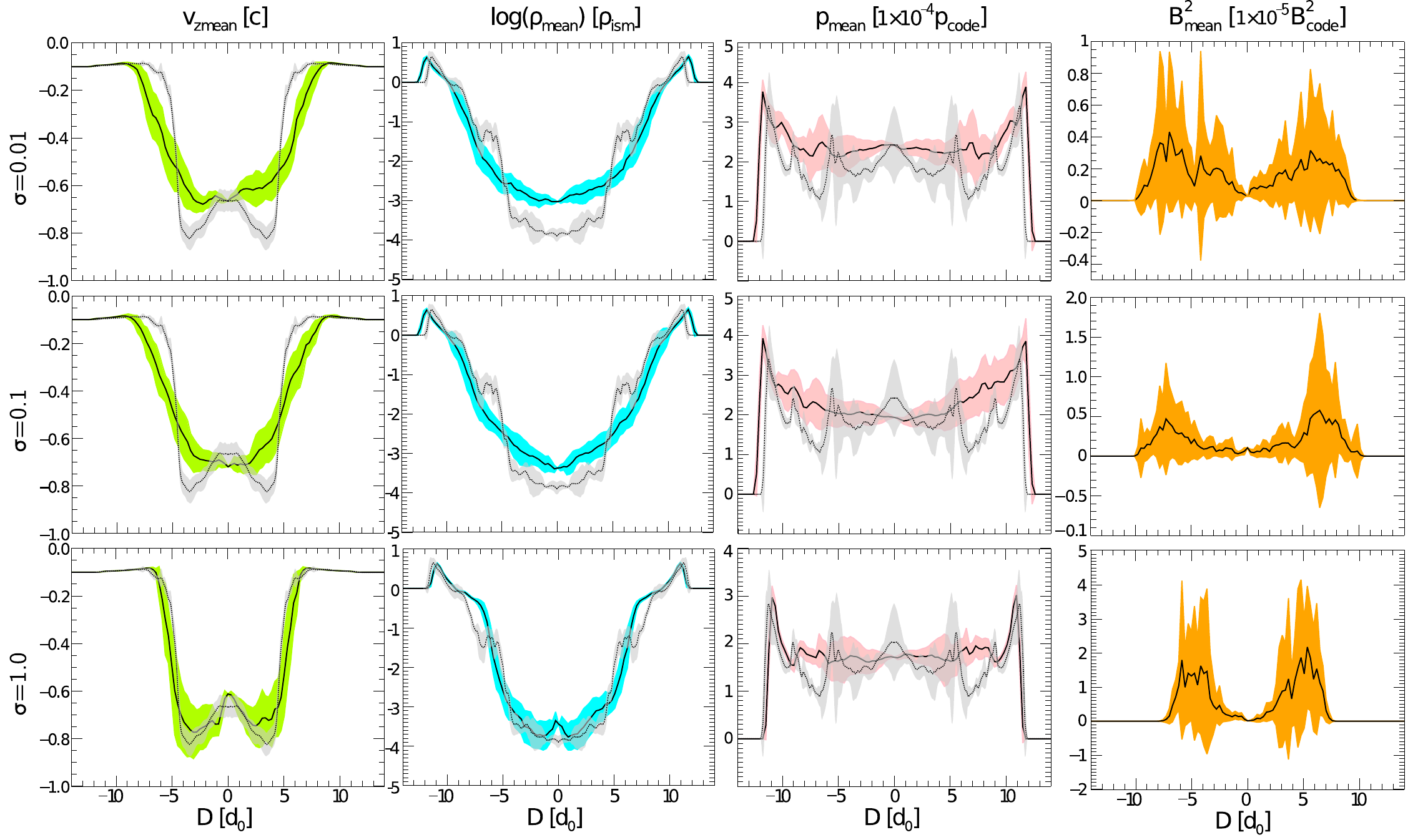}
   	 \caption{Same as in Fig.~\ref{fig:cfr_B0_ISO} but for run A$_{\{0,i\}}$. Again black solid lines indicate the computed average trend of each variable and the surrounding colored area the related standard deviation; black dotted lines and gray areas represent trends extracted from run 2D-H$_0$.}
	\label{fig:cfr_B0}
   	 \end{subfigure}\hfill	
\end{figure*}

To characterize the differences in the flow structure between the 2D and 3D cases, we have looked at the behavior of various fluid and magnetic quantities along the tail, performing transversal cuts. These have been extracted as 2D slices in the $(x,\,y)$ plane of the 3D data cube, at three distinct positions along the bow shock tail: $z=[-20.0,\,-20.8,\,-21.7]d_0$. 
This choice has been done for two reasons: at these locations the tail has reached a quasi-stationary configuration, and they are far enough from the bow shock head that one can neglect to first order the dynamics of the very head;  on the other hand the region very close to the boundary of the numerical grid has been excluded to avoid numerical artifacts. 
We sample every dynamic variable along different transverse directions, that can be though of as diameters of an arbitrary circumference of fixed radius ($\sim13 d_0$),  centered on the symmetry axis ($x=y=0$), as illustrated in Fig.~\ref{fig:LinOuts}, where they are drawn as colored lines.
Profiles are given as functions of the position along these directions, and the directions indicated by arrows in Fig.~\ref{fig:LinOuts}.  In this representation the symmetry axis is located at $\mathcal{D}=0d_0$ and the plane containing the pulsar spin axis is the one selected by the vertical line identified by label $\mathcal{L}$.
The profiles obtained along each of these directions are then averaged. We first average at fixed $z$, spanning all the different diameters, and then average over the three different slices (thus varying $z$). The variability is then encoded by the standard deviation, which is also shown (in grey for 2D and different colors for 3D). By the way we perform our averages, the standard deviation has a minimum at the center of the circumference. 
Profiles obtained with this procedure are then compared with similar profiles obtained from the 2D simulation 2D-H$_0$. 
In this case we have used  six cuts in the same range $z=[-20.0;-21.7]d_0$ at three consecutive time-steps, to derive more robust averages.
Results are shown in Fig.~\ref{fig:cfr_B0_ISO} and Fig.~\ref{fig:cfr_B0}. 

We found that the mean properties of the tails show very similar trends in all cases, thus we only show few representative runs, focusing on the role of the magnetic field: I$_{\{0,i\}}$  and A$_{\{0,i\}}$, both with $\phi_M=0$. As we will discuss in the following,  deviations from these trends are seen only for runs with $\phi_M=\pi/4$.
 
%
From a first look at  Fig.~\ref{fig:cfr_B0_ISO} and Fig.~\ref{fig:cfr_B0} no major asymmetry is observed, consistent with the axisymmetric injection of these runs.
As discussed before the major difference  is  the lateral extension of the tail marked by the position of  the contact discontinuity. In the 2D hydrodynamic case its diameter is of the order of $\sim 10d_0$ on average. In the 3D cases instead the extent of the tail is larger  for lower values of the magnetization ($\sim 13d_0$), and only in the case $\sigma=1.0$ we recover a size comparable with the 2D-H$_0$ case. What we see here is that the density jump at the CD is shallower al lower magnetization.  
The extension and variability of the shear layer can be estimated from Figs.~\ref{fig:cfr_B0_ISO}-\ref{fig:cfr_B0}, looking to the $v_z$ plots. 
It can be seen that the shear layer become less extended while increasing magnetization for both the isotropic and anisotropic models, going from $\sim 6.5d_0$ for $\sigma=0.01$ to $\sim 3d_0$ for $\sigma=1.0$, with a maximum variability of $\sim0.9d_0$. 
These number must be compared with the 2D case, where the shear layer extension is of order $\sim4d_0$, with a small variability of $\sim0.2d_0$. 
As already pointed out before, as the magnetization increases the 2D prediction is recovered.

The properties of the shear layer can be equivalently extrapolated from density plots. In the 2D case there is a sharp jump between the PWN material and the shocked ISM.
On the contrary this contrast is much less evident in the 3D cases. Here the higher value of the density in the center is a clear signature of the mixing with the denser ambient medium, which is more efficient in 3D than in 2D.
The only exception is seen for case I$_{\{0,\,1\}}$, which was already indicated as the most similar to the 2D one. In that case the density within the PWN is compatible with the one found in 2D.
In any case the inner density approaches the 2D one with increasing magnetization, meaning again that the importance of the turbulence diminishes with increasing the magnetization level.

This confirms what was seen before in term of turbulence developing at the shear layer between the PSR material and the shocked ISM. Interestingly for models with an anisotropic energy injection the tail appears to be more extended than in the isotropic ones. 
This can be again explained in terms of turbulence: anisotropic energy injection tends to create stronger shear layers in the very head, and a more turbulent flow is injected from the head of the system. 
%

\begin{figure*}
	\centering
	\includegraphics[width=.495\textwidth]{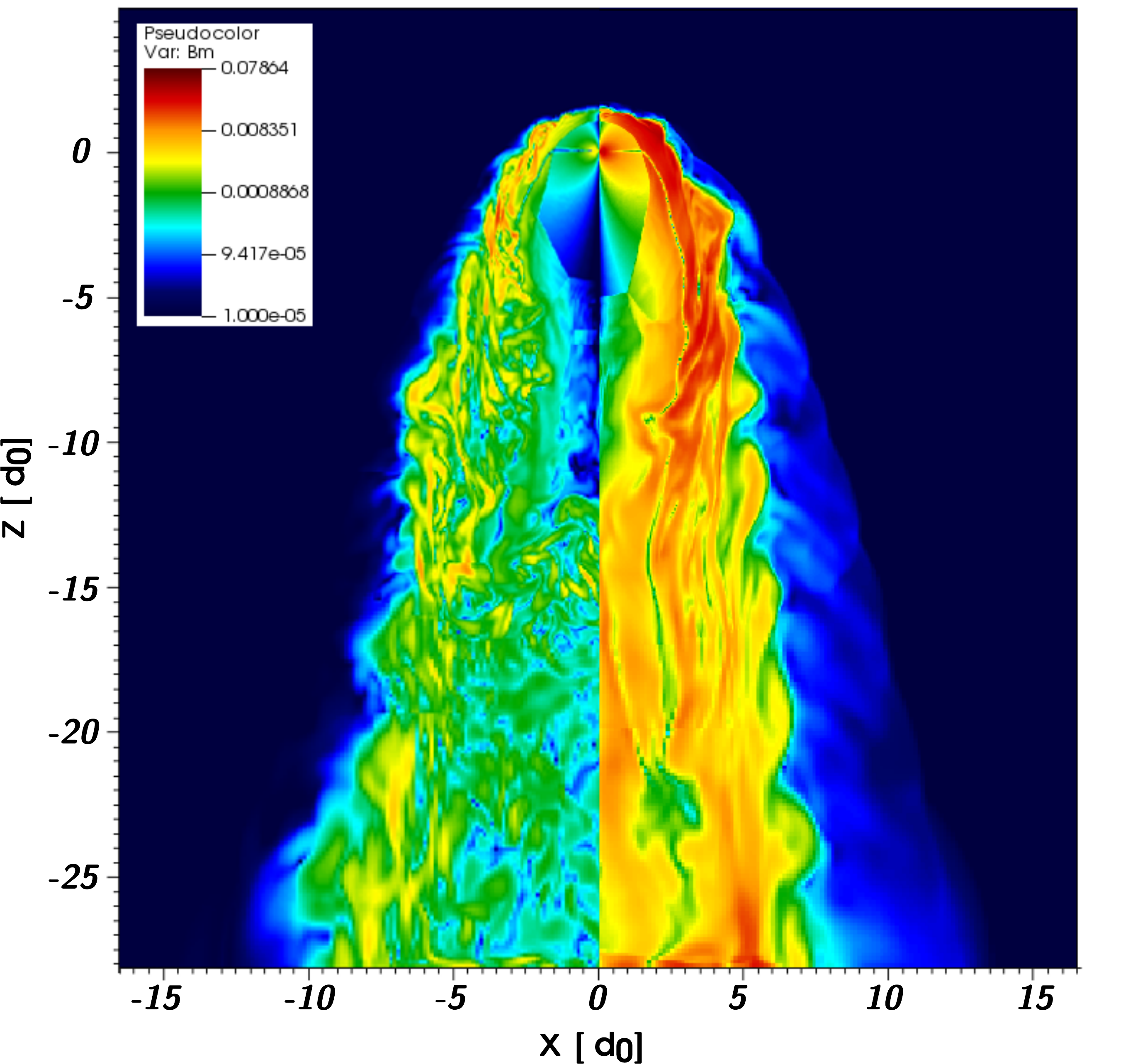}\,
	\hspace{-0.5cm}
	\includegraphics[width=.495\textwidth]{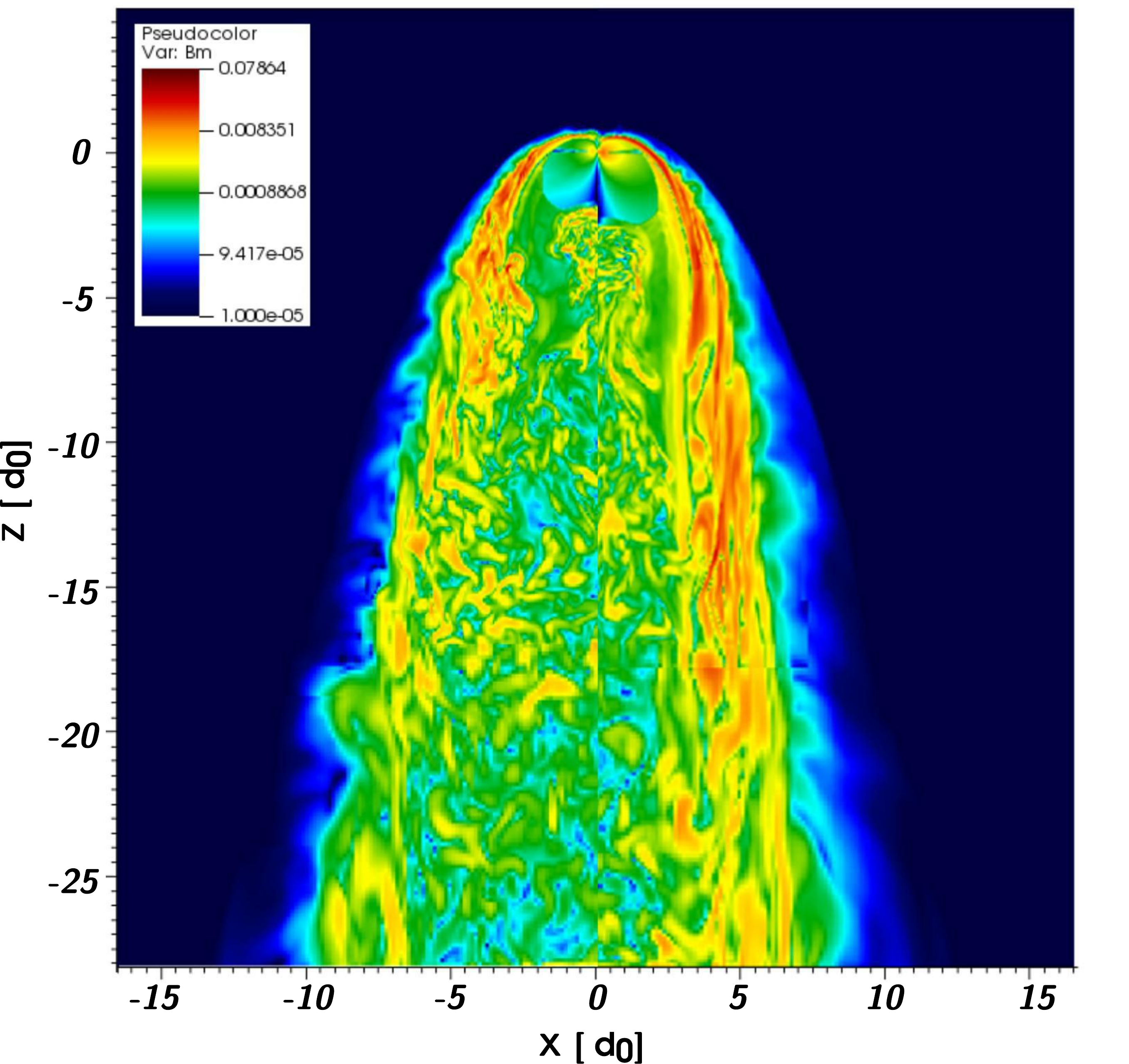}\\
	\vspace{-0.0cm}
	\includegraphics[width=.999\textwidth]{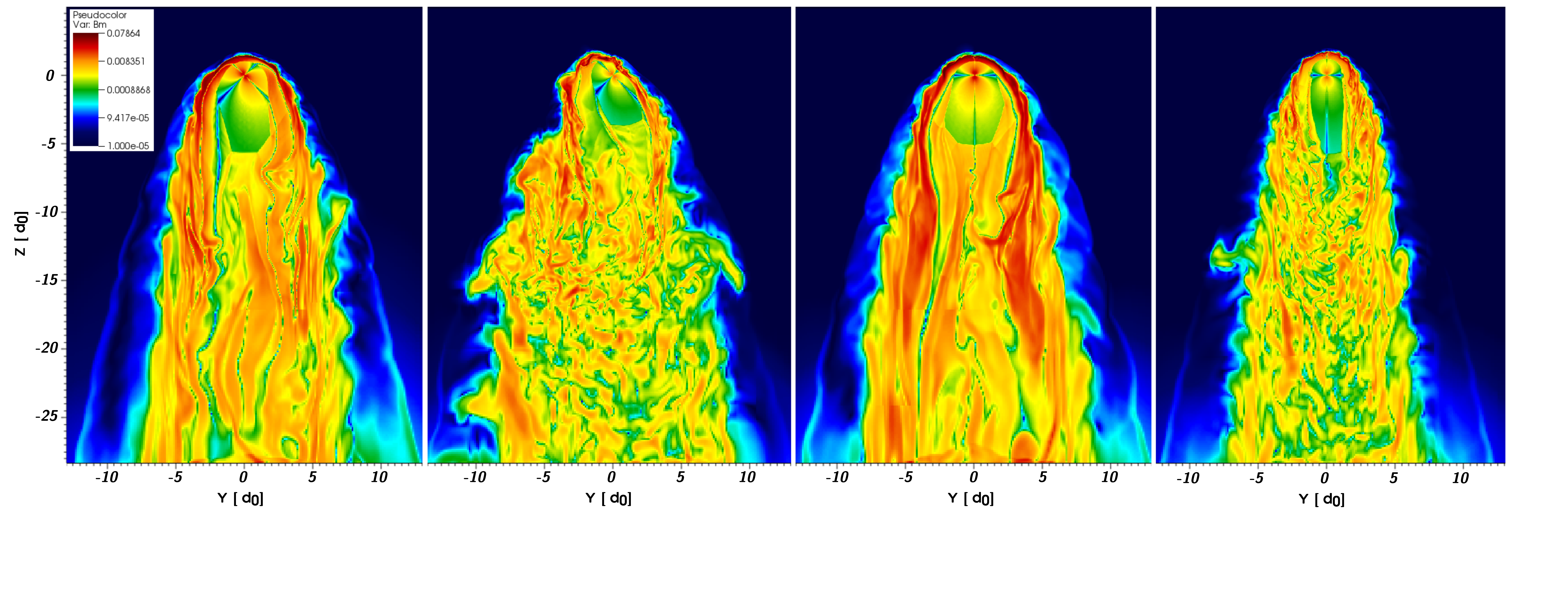}\\	
	\vspace{-1.0cm}
	\caption{Map of the magnetic field module (in code units), normalized to the maximal nominal value. Upper panel: comparison for runs I$_{\{ 0,i\}}$ (on the left) and A$_{\{ 0,i\}}$ (on the right); maps are composed by two halves of different value of magnetization, $\sigma=0.01$ on the left-side and $\sigma=1.0$ on the right-side, respectively.
	Bottom panel: maps of the magnetic field module for magnetization $\sigma=1.0$ and for the all the remaining inclinations $\phi_M$. From left to right maps refers to run:  I$_{\{ \upi/4,0\}}$, A$_{\{ \upi/4,0\}}$, I$_{\{ \upi/2,0\}}$ and A$_{\{ \upi/2,0\}}$. 
	Please notice that the color map is the same for all plots.
	}
	\label{fig:cfrB}
\end{figure*}

The third columns of Fig.~\ref{fig:cfr_B0_ISO} and Fig.~\ref{fig:cfr_B0} show the thermal pressure inside the bow shock nebulae. A common feature for both the isotropic and anisotropic cases is the decrease of the average pressure with increasing magnetization (as the consequence of the rising of the magnetic pressure). 
For $\sigma=0.01$ pressure is almost everywhere higher than the 2D case, while for $\sigma=1.0$ the 3D trend is more or less the same than the 2D one. 
In all cases pressure appears to be highly variable in the tail, with the maximum reached at the FS surface. 
It is also interesting to notice that in 2D the pressure shows a rapid drop after the CD, at $D=\pm 7d_0$, which is not shown by any of the 3D runs, where pressure is maintained quite uniform up to the boundary or it even presents an increase at the same location.
Moreover pressure plots do not show evident differences between the isotropic and anisotropic models.

The difference between the isotropic and anisotropic cases is again visible when comparing the velocity profiles.
Going back to Fig.~\ref{fig:cfr_B0_ISO}-\ref{fig:cfr_B0}, the first columns displays the $z$ component of the velocity, aligned with the pulsar motion. 
In the isotropic models the velocity shows a slower channel around the symmetry axis similar to what observed in 2D, even if it is not as marked, where velocity is, in absolute value, of order of $\sim 0.65c$ within a distance of $\sim0.8d_0$, to be compared with the fast outer channel with $\abs{v_z}\sim 0.82c$. 
In the 3D isotropic cases the maximum velocity variation from the inner to the outer channels is seen for case I$_{\{0,\,1\}}$, where velocity goes from $v_z\sim 0.7c$ to $v_z\sim 0.82c$ and the extension of the slow velocity channel is of order of $d_0$ in radius.
This behavior is however absent for models with  anisotropic energy injection where, on the contrary, velocity is usually maximum along the $z-$axis. 
Again this is an effect of different level of turbulence that can destroy the coherence of flow structures. Moreover in the isotropic models $v_z$ remains higher if compared with the anisotropic runs. In the isotropic cases the average speed within a region of extension $\sim5 d_0$ around the symmetry axis is $\abs{v_z}\gtrsim0.75c$, and the flow appears to be more collimated around the $z$ axis.
On the contrary in the anisotropic case the velocity is slightly lower (with $ 0.6c \lesssim \abs{v_z}\lesssim 0.8c$) and the high velocity flow is distributed on greater distances from the symmetry axis, following the broadening of the tail.

The different properties of the tail are also influenced by the level of magnetization that survives in the tail and by the structure of the local magnetic field, which play an important role on the development of the turbulence.
In Fig.~\ref{fig:cfrB}  the magnitude and morphology of the magnetic field are shown for some reference runs at $t=t_f$.  
The upper panel compares the field intensity and morphology for the minimum ($\sigma=0.01$, leftmost part of each map) and maximum ($\sigma=1.0$, rightmost part of each map) values of initial magnetization for the aligned isotropic run (i.e. I$_{\{ 0,\, i\}}$, left panel) and the anisotropic one (i.e. A$_{\{ 0,\, i\}}$, panel on the right).
The bottom panel of the same figure shows the whole map of the magnetic field for the same magnetizations for both the isotropic and anisotropic models and magnetic inclinations $\phi_M=\upi/4$, $\phi_M=\upi/2$.
Here it is evident how the survival of highly magnetized regions far from the bow shock head helps in maintaining coherent structures on large spatial scales and that injection with a higher magnetization (left-half part of each map) prevents the formation of small scale turbulence. 

The comparison between different values of the magnetization shown in the upper panel of Fig.~\ref{fig:cfrB} for runs I$_{\{0,i\}}$ and A$_{\{0,i\}}$, illustrates that in the case of low magnetization the magnetic field is almost fully dissipated and randomized already at distances of the order of $z\sim 15 d_0$ from the pulsar. 
On the contrary for $\sigma=1.0$ the magnetic field survives far further from the BSPWN head: the random motion in the tail is lower and coherent structures are clearly visible on large scales and the tail is maintained more collimated around the symmetry axis.
The global maximum of the magnetic field is not so different from case to case and from model to model, with the maximum usually arising near to the front shock at the BSPWN head, where the injected magnetization is still higher (due to the minor dissipation it has suffered) and the field is enhanced by shear  instability and compression. 
Values of the maximum ranges between $2\times10^{-2}$ and $\sim10^{-1}$ in code units, corresponding to $\sim 10 - 50\, (\rho_{24} v_7^2)^{1/2}\,\mu$G. 
The global average value of the field is of the order of $10^{-3}$ in code units or $0.5 \, (\rho_{24} v_7^2)^{1/2}\,\mu$G, as can be seen form the extracted average value in the tail.

Due to turbulence and dissipation the average value of the magnetization is always less than the injected value, with values of a few$\times 10^{-4}$ for $\sigma=0.01-0.1$ up to a few$\times10^{-3}$ for $\sigma=1.0$.
The same behavior is shown from runs with different inclinations of the magnetic field (bottom panel of the same figure).

Compared to other variables as density and velocity, the magnetic field appears also to have the higher level of variability, as can be seen looking at the standard deviations in Fig.~\ref{fig:cfr_B0_ISO}-\ref{fig:cfr_B0}. 
Interestingly the magnetic field seems to reach the maximum near to the contact discontinuity, possibly reflecting the presence of an effective amplification mediated by the shear instability, even if in the low magnetized regime, early 2D simulations were suggesting also the possibility of local compression into a magnetopause \citep{Bucciantini:2005}.
On the other hand for higher magnetization we see a local enhancement of the magnetic field strength close to the axis. 
This is reminiscent of results in 2D RMHD, and it is due to a residual effect of hoop stresses associated with the toroidal component of the magnetic field, that tends to pinch the flow on the axis. However the effect is reduced in 3D, where there are no symmetry constraints.

\begin{figure*}
	\centering
	\includegraphics[width=.95\textwidth]{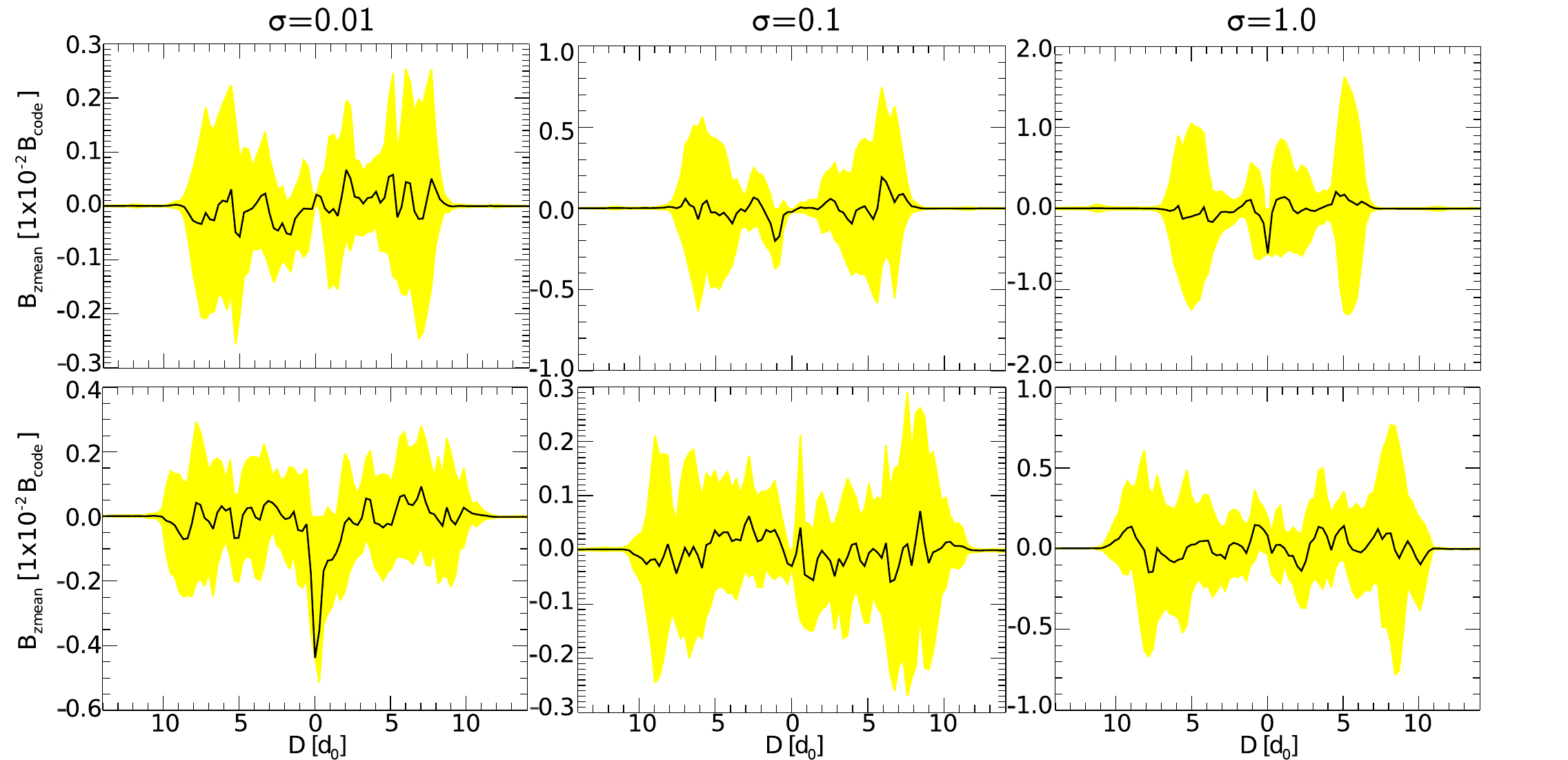}\\
	\caption{1D profiles for the $B_z$ component of the magnetic field as function of the distance along the diameter $\mathcal{D}$, in units of $d_0$. 		Upper panel shows plots from runs I$_{\{\upi/4,i\}}$, bottom for A$_{\{\upi/4,i\}}$.
	Solid black lines indicate the average value of $B_z$ and yellow color areas the corresponding standard deviation. 
	The magnetic field component is here given in $10^{-2} B_\mathrm{code}$ units.
	}
	\label{fig:BzCFR}
\end{figure*}

A closer look to the magnetic field variations can help in investigating the nature of the turbulence. 
In Fig.~\ref{fig:BzCFR}, 1D profiles for the $B_z$ component of the magnetic field are given, for all the magnetizations, considering cases I$_{\{\upi/4,i\}}$ and A$_{\{\upi/4,i\}}$, for which the effect of turbulence appear stronger. 
Please beware that the scales of the plots, given in units of $10^{-2}B-\mathrm{code}$, are not the same for the different plots, since these have been chosen in order to highlight the field oscillations. Differences in the amplitude must be then read taking into account the different scales.
First thing to notice is that the magnetic field has an intrinsic turbulent nature, clearly visible from the fact that the average value of $B_z$ is compatibile with 0, with very high oscillations.
%
In the case of the anisotropic wind (bottom row) for $\sigma\le 0.1$ the increase of the injected magnetization does not reflect in a growth of the magnetic field: maximum values and amplitude of the variability do not change, with the maximum approaching 0.002 in code units, i.e. $1 \, (\rho_{24} v_7^2)^{1/2}\,\mu$G. 
This means that in this regimen the magnetic field evolution and dynamics is actually dominated by turbulence rather than injection. 

On the contrary, as $\sigma$ exceeds 0.1, the magnetic field is seen to rise with increasing initial magnetization, even if not linearly (juts a factor $2$ for a rise of magnetization by a factor 10), meaning that the system now retains memory of the injection. 

In the isotropic wind model (upper row), each step in magnetization corresponds to an increment of a factor of 2 in the maximum value of $B_z$, which goes approximately from 
$0.002$ to $0.1$ in code units, i.e. from $1(\rho_{24} v_7^2)^{1/2}\,\mu$G up to $50 (\rho_{24} v_7^2)^{1/2}\,\mu$G.
In this case injection is evidently dominant and the development of small scale turbulence is less efficient.

\begin{figure*}
	\centering
	\includegraphics[width=.95\textwidth]{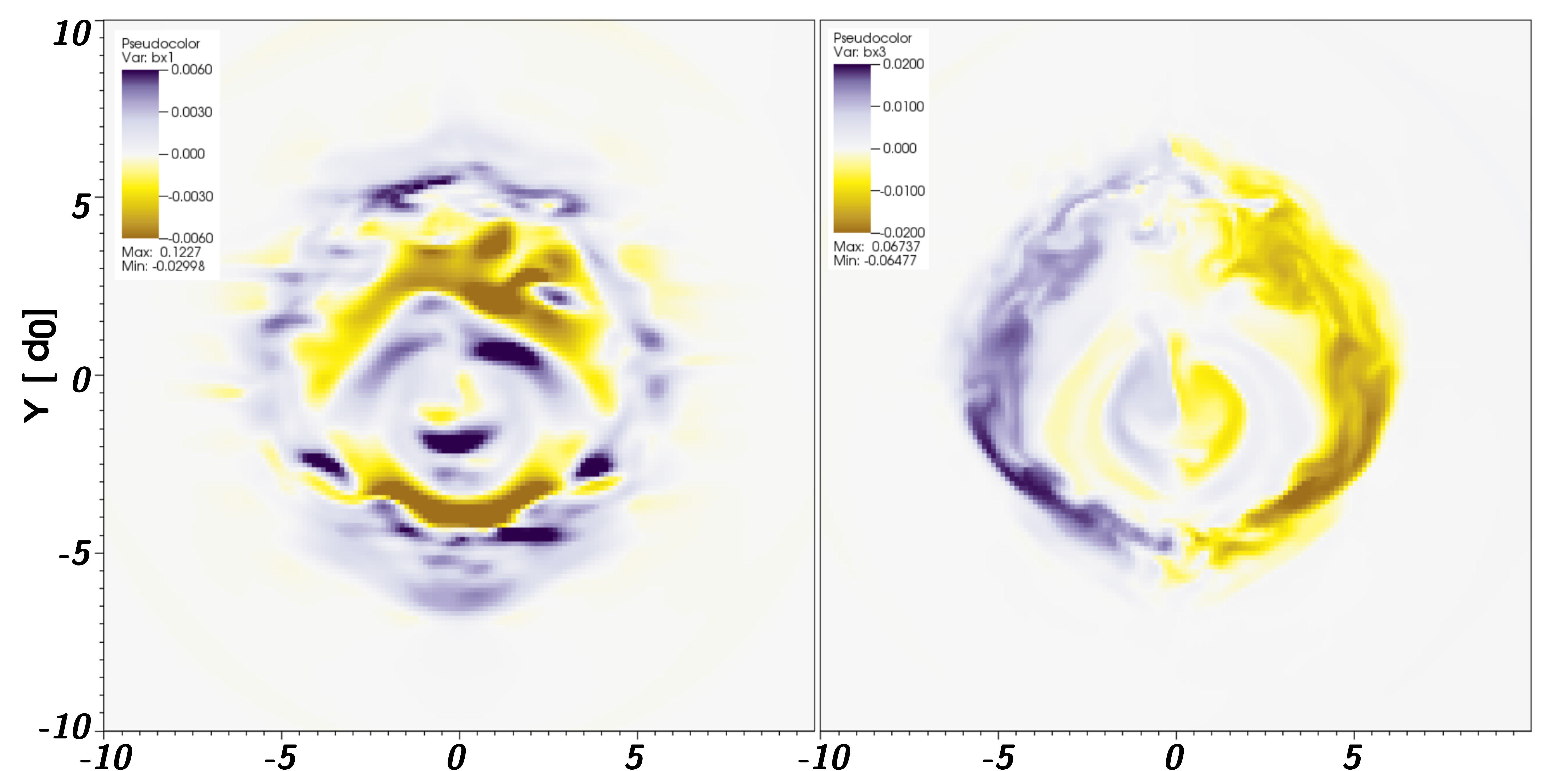}\\
	\vspace{-0.01cm}
	\includegraphics[width=.95\textwidth]{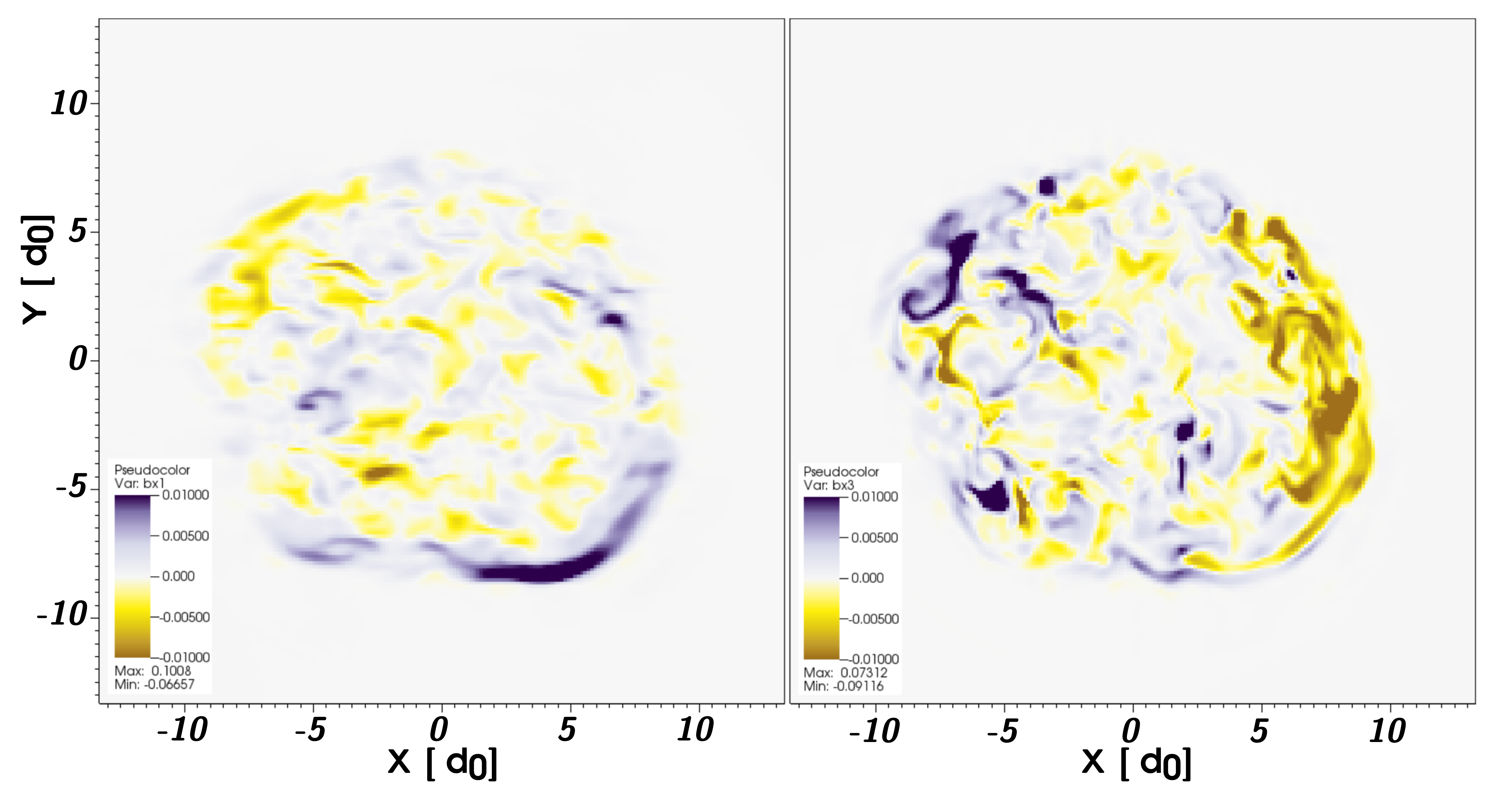}\\
	\caption{2D maps of the magnetic field components $B_x$ (panels on the left) and $B_z$ (panels on the right), obtained as $(x,y)$-slices in the tail at $z=-20.8$. Upper panels show case  I$_{\{\upi/4,0\}}$ while bottom panels show case A$_{\{\upi/4,0\}}$. 
	Please notice that, in order to make small structures more visibile, the spatial scales are different from upper to bottom panel. 
	The intensity of the magnetic field components is given by the color scale in the bottom-left part of each map, with different ranges: upper-left $Bx\in[-0.006,\,0.006]\, B_\mathrm{code}$, upper-right $Bz\in[-0.02,\,0.02]\, B_\mathrm{code}$, bottom-left $Bx\in[-0.01,\,0.01]\, B_\mathrm{code}$, bottom-right $Bz\in[-0.01,\,0.01]\, B_\mathrm{code}$.
	}
	\label{fig:TURB}
\end{figure*}

Planar components of the magnetic field can be seen in Fig.~\ref{fig:TURB}, where $B_x$ and $B_z$ are plotted as 2D $(x,y)$ slices (at $z=-20.8d_0$) for runs I$_{\{\upi/4,0\}}$ and A$_{\{\upi/4,0\}}$. 
The nature of turbulence appears very different between the isotropic and anisotropic wind models , with the anisotropic case dominated by small scale structures.
Again these differences causes clear diversities in the field morphology: in the isotropic case the field shows coherent structures on large scales, and the injection geometry is still recognizable in the tail far away from the head, as can be seen by the different polarities of the field still well separated (yellow color vs violet).
On the contrary in the anisotropic model the high level of mixing causes this global structure to be very poorly recognizable in the tail. The $B_z$ component maintains a very faint left-right separation of polarities, while the mixing in the $B_x$ component (and $B_y$) is almost complete, and no coherent structure is visible any more.
This behavior is common for almost all the considered configurations. 
The exception, as already pointed out in the previous discussion, is represented by the low magnetized cases ($\sigma=0.01$), where the level of turbulence is high enough to dominate on the injection properties and to destroy the large scale structure of the field also in the isotropic case. 
In the isotropic case the planar component of the magnetic field peaks in a region extending between $1-4d_0$, and it shows a shell-like morphology (with thickness of order of $\sim 1.5d_0$), with subsequent well separated shells of different polarities. 
Here the maximum value is of the order of $\pm 5 (\rho_{24} v_7^2)^{1/2}\,\mu$G. 
On the contrary the $z$ component peaks near to the CD in a shell of variable thickness  ($\sim1.3d_0$ at the bottom up to $\sim 2.5d_0$). The magnetic field is particularly intensified  in the bottom region, $y<0$, showing a magnification of $\sim 3.5$, while it shows a minor enhancement (a factor of $\sim1.5$) for $y>0$. This enhancement is probably the effect of shear instability acting at the CD surface.
On the other hand in the anisotropic case both the planar and $z$ components of the magnetic field peak near to the CD, in different regions of the $(x,\,y)$ plane. Moreover the maximum value is more or less the same, still indicating that the magnetic field is mostly dominated by turbulence in all the directions.
%
\begin{figure}
	\centering
	\includegraphics[width=.5\textwidth]{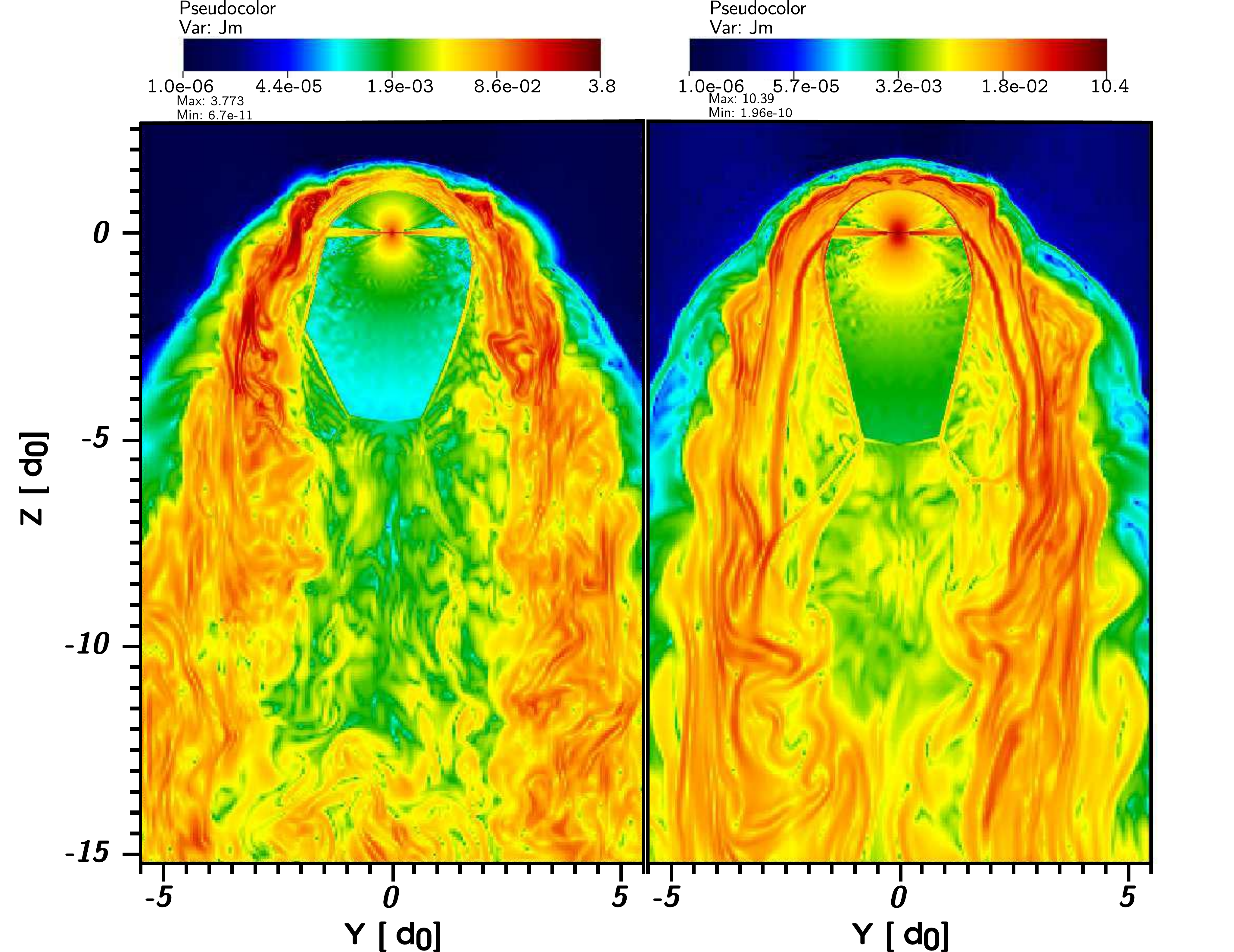}
	\caption{2D slices  color maps in the head region of case I$_{\{ 0,2\}}$ (left) and I$_{\{ 0,0\}}$ (right) of the current density given by $(\mathbf{\nabla} \times \mathbf{B})$ (the displacement current $\partial \mathbf{E}/\partial t$ is negligible in all the cases).	}
	\label{fig:CS}
\end{figure}

A close-up view of the bow shock head is given in Fig.~\ref{fig:CS}.
Here the head morphology is compared for cases I$_{\{ 0,2\}}$ and  I$_{\{ 0,0\}}$ as 2D maps of  the module of the current density, given in code units. 
This zoom-in helps to see that the pulsar wind region is perfectly resolved in our simulations, even for the highly magnetized case (where the shock is expected to be smaller). 
The injection region, i.e. a sphere of radius $\sim 0.2d_0$ centered on the pulsar, appears in fact well far away the TS surface in each direction. 
From this figure the evolution of the current sheet can be followed, comparing low and high-$\sigma$ cases. 
The current sheet originates at the pulsar equatorial plane and appears to be evidently different in the two cases. In the low$-\sigma$ case it maintains its structure up to a distance of $\sim 5d_0$ from the pulsar. At that point vortexes start to mix the current sheet with the surroundings, causing it to twist and tangles while it is completely mixed up. 
At a distance of $\sim 12d_0$ from the pulsar, the current sheet is no  more recognizable from the background.
On the other hand, when magnetization is higher, the current sheet survives and it is easily visible up to $\sim 14d_0$ from its birth region. 
It maintains a quasi-laminar structure up to a distance $\sim 10 d_0$ from the pulsar, then it starts to be diverted.

Again these differences points out that in presence of an higher magnetized flow turbulence is less efficient in mixing up the plasma and quasi-laminar structures can survive far away from the bow shock head.
The high$-\sigma$ case also shows a maximum value of the current greater than a factor of $\sim 3$ with respect to the low magnetized case.
When magnetization is higher the current also appears to be enhanced at the CD surface.
 Notice that the high-magnetized case is also the one in which the magnetic field feels the major increment as the consequence of the rise in $\sigma$.

%
\begin{figure}
	\centering
	\includegraphics[width=.48\textwidth]{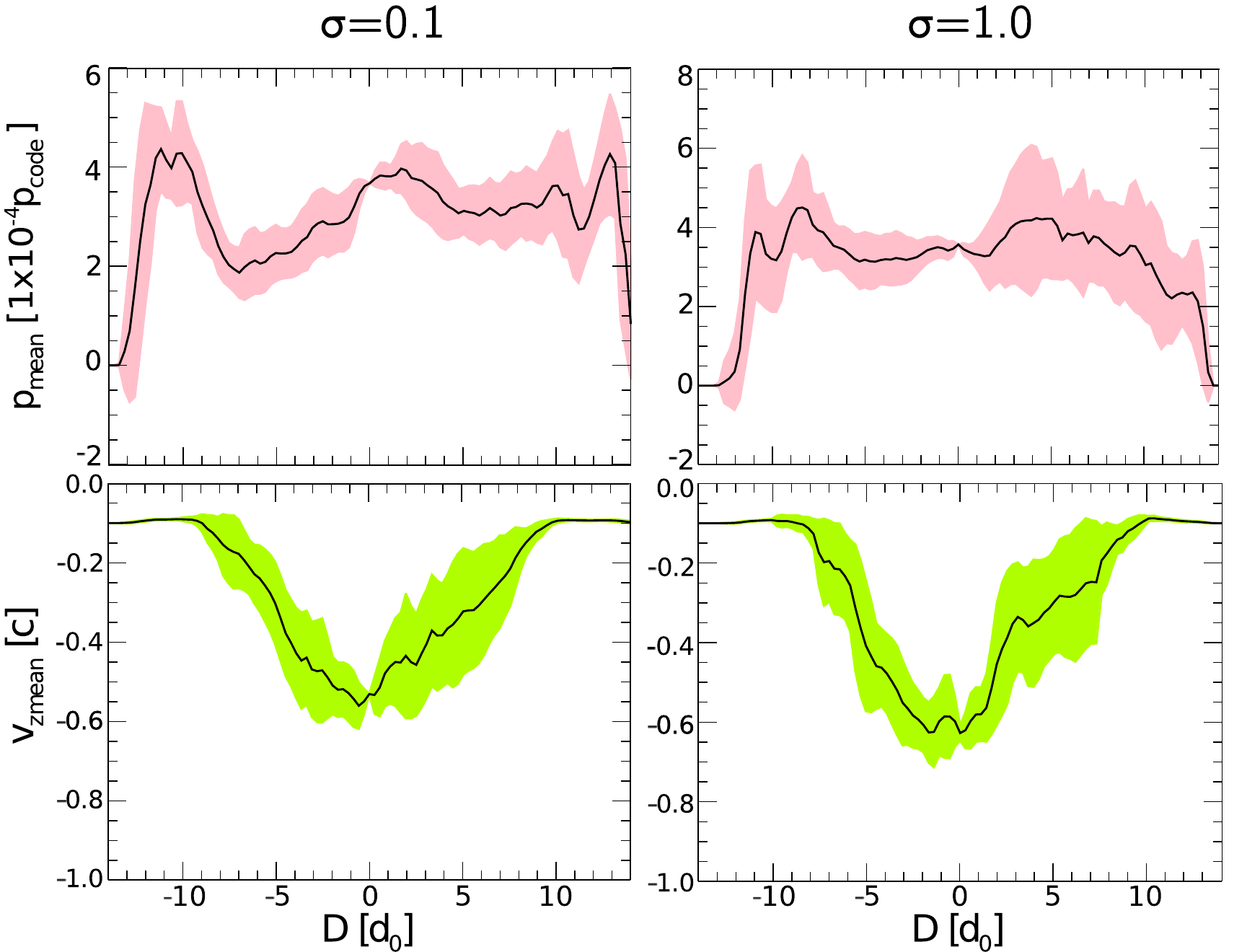}	
	\caption{1D average profiles for pressure (upper row, in units of $10^{-4}p_\mathrm{code}$ units) and $v_z$ (bottom row, in units of $c$), referring to cases A$_{\{ \upi/4,1\}}$, left panel,  and A$_{\{ \upi/4,0\}}$ panel on the right.
	}
	\label{fig:cfrB45}
\end{figure}

\begin{figure}
	\centering
	\includegraphics[width=.49\textwidth]{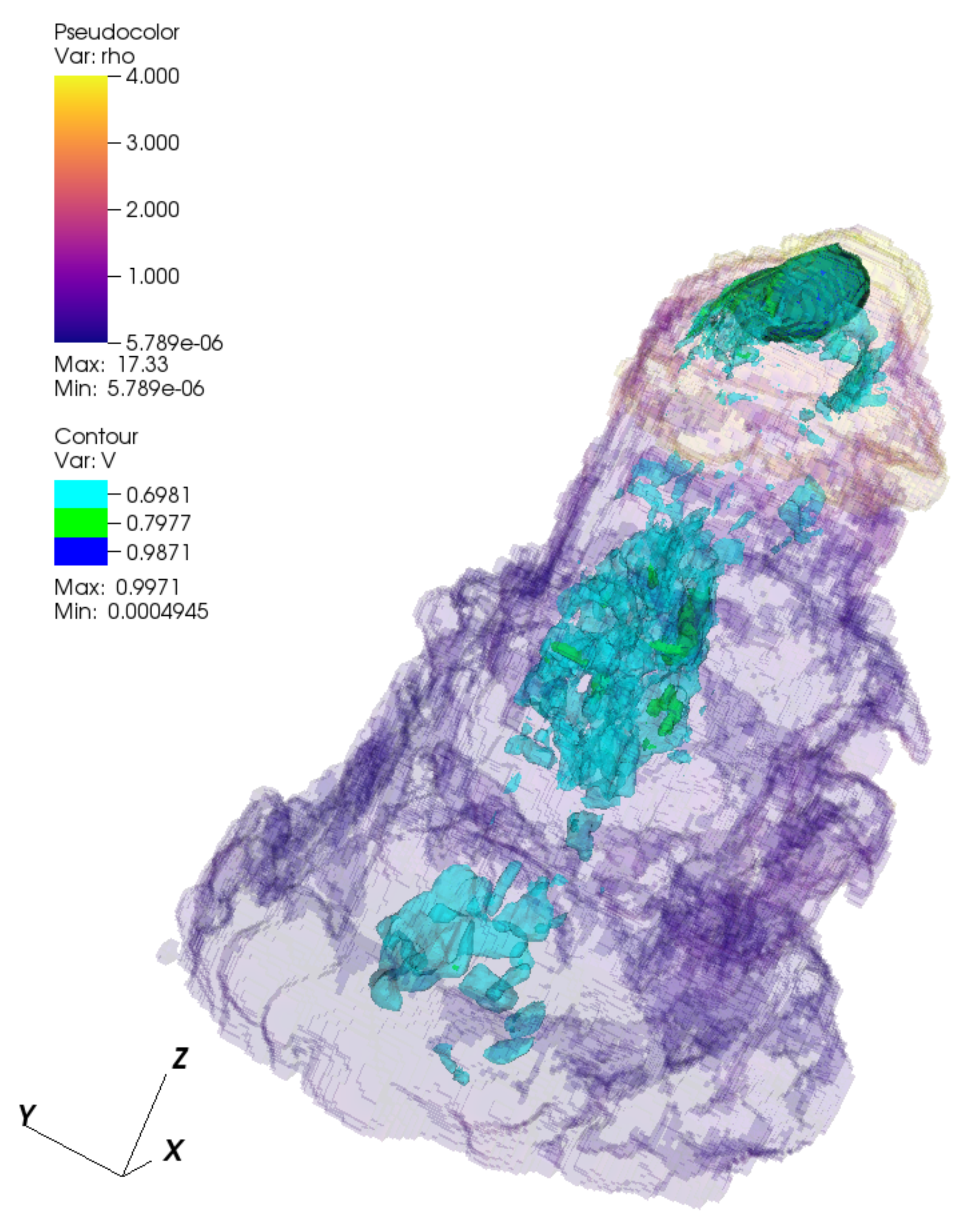}	
	\caption{3D composite map of the density and velocity magnitude for run A$_{\{ 0,1\}}$, at $t=t_f$. The density map is made with  transparent isocontours at ten different levels along the logarithmic color scale, with uniform spacing. Velocity is shown at the three fixed values  shown in the legend and contours are made transparent in order to appreciate the third dimension.
	}
	\label{fig:B45rho}
\end{figure}

As anticipated major deviations from the common trends shown by the average quantities in the tail arise in cases with $\phi_M=\upi/4$.  This can be observed in Fig.~\ref{fig:cfrB45}, where profiles for pressure and $v_z$ are shown.
They present a clear left-right asymmetry, which gives rise to a distorted shape of the BSPWN with respect to the other inclinations. 
The global structure of the bow shock is visible in Fig.~\ref{fig:B45rho}, where isocontours of the velocity magnitude are superimposed to a 3D map of density for run A$_{\{ 0,1\}}$, to highlight the asymmetry of this configuration with respect to the $z-$axis. 
As predicted the TS has an oblate shape, with major extension in the direction inclined by $\upi/4$ with respect to the $z-$axis. 
Velocity is only shown for values $\gtrsim 0.7c$ and only a few spots of $v\sim 0.8c$ are visibile in the tail, concentrated at approximately the half of the tail extension and in the left-side part of the image, with Lorentz factor $\sim1.6$.
This asymmetry, which also reflects in a global asymmetry of the entire bow shock, is a direct effect of the shock shape and inclination. 
In this configuration the flow coming from the TS has very different evolution depending on the part of the shock at which it is originated: when the elongated  part of the TS points to the outer shock (in the right-hand part of the figure, pointing in the $+z$ direction), the flow is directed towards the outer shock, where it is diverted and decelerated.
On the contrary, when the elongated part of the shock points to the back of the tail, in the $-z$ direction, the high velocity flow coming from the shock is not decelerated by the interaction with the outer shock and it may survive along the tail.
This is in fact what is seen in Fig.~\ref{fig:cfrB45}, where almost all the high velocity flow ($v>0.7c$) is located behind the part of the shock pointing in the $-z$ direction rather than in the opposite.
This obviously leads to an asymmetric shape of the entire object and produces a preferential region with high velocity in the direction in which the alignment of the shock front with the pulsar motion is maximized.
For obvious geometrical reasons the same effect is not seen when $\phi_M=\upi/2$ or $\phi_M=0$.

\begin{figure}
	\centering
	\includegraphics[width=.49\textwidth]{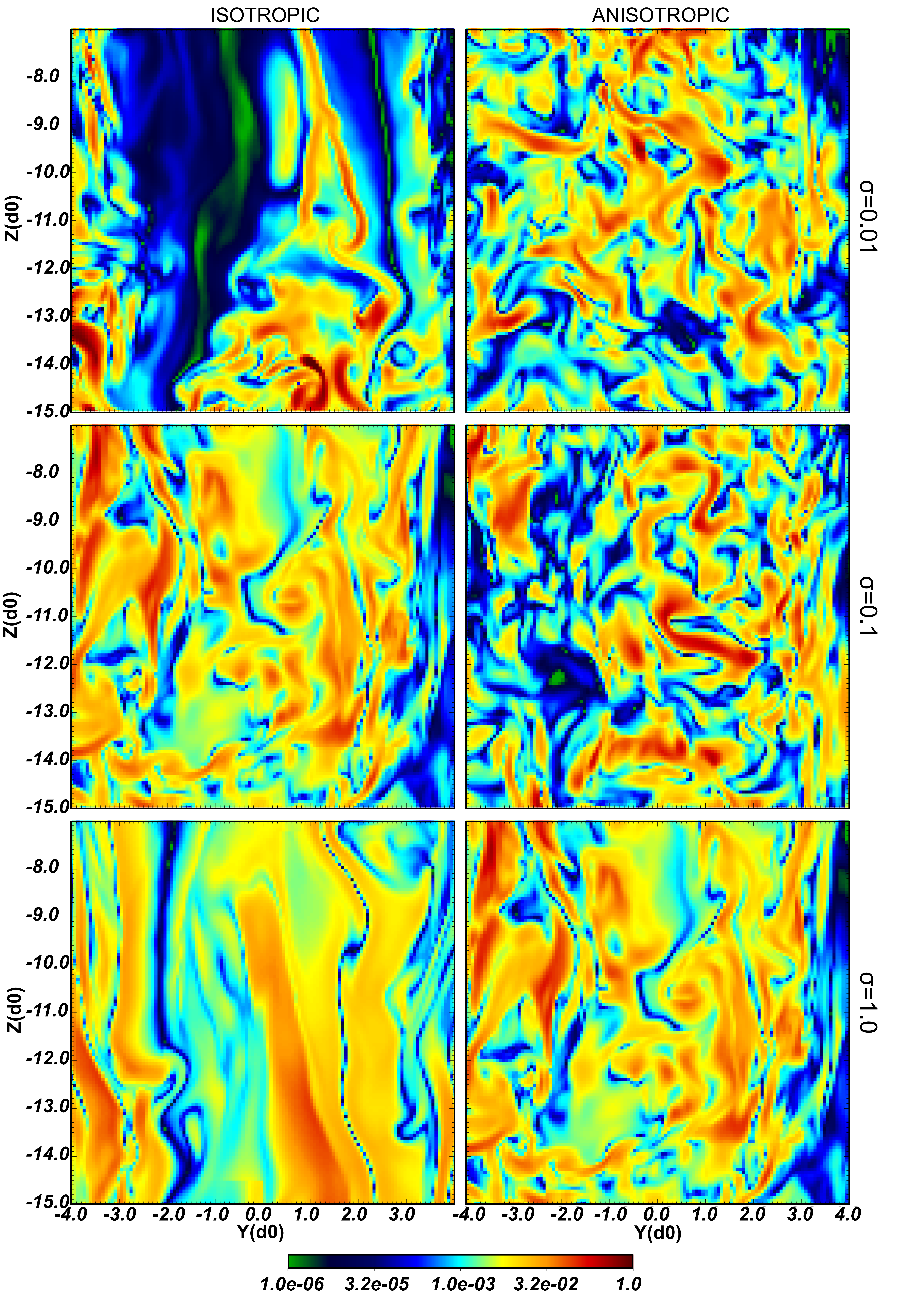}	
	\caption{2D slices  color images of the magnetization in the tail, with the appropriate definition for a relativistic gas $B^2/(4p +\rho)$, at $t=t_f$, in the region in the bow shock tail defined by $z\in[-15,-7]d_0$, $y\in[-4,4]d_0$, and $x=0$. Maps are given for all the models with $\phi_M=\upi/4$. Left-side column shows images for the isotropic models, right-side column the anisotropic models. Magnetization changes from top to bottom. Maps are normalized to their respective maxima: from the lowest to the highest $\sigma$ 0.3, 0.9, 1.8 for the isotropic cases, and 0.8, 0.3, 1.6 for the anisotropic ones.}
\label{fig:turbC}
\end{figure}

A closer look to the local structure of the flow in the bow shock tails for all the magnetic configurations for  $\phi_M=\upi/4$ can be seen in Fig.~\ref{fig:turbC} . 
Here 2D $(y,z)$ maps of the magnetization are shown to compare scales of the turbulence in the isotropic (left column) and anisotropic (column on the right) models.
Maps are normalized to their own maxima and only shown in a region of the tail identified by $z\in[-15,-7]d_0$ and $y\in[-4,4]d_0$. 
Even at a local view the different dynamic of the shocked wind between the isotropic and anisotropic models is  evident, especially when increasing the initial magnetization. 
In the anisotropic model small scales turbulence is dominant at all the magnetizations, with the smallest eddies showing a dimension of $\sim 0.3d_0$ and being dominant in the flow structure for the lower magnetized cases.  
On the contrary in the isotropic model small scales turbulence is only poorly dominant and disappears when the initial magnetization is increased. 
As already discusses the case with $\sigma=1.0$ shows a very ordered flow dynamics, with a quasi-laminar appearance, with elongated structures well recognizable along the pulsar direction of motion. 

Among all our runs, case A$_{\{\pi/4,2\}}$ is the one showing the more developed turbulence in the velocity field. In order to characterize this turbulence we have  computed the second order structure function of the velocity. Traditionally the structure function is computed for the longitudinal and trasverse components (with respect to the displacement) of the velocity.  Here we instead compute the second order  structure function of velocity components $v_x$ and $v_y$ perpendicular to the axis of the tail, in planes taken at different locations in $z$. We do this for several reasons: first $v_z$ in the tail shows a clear pattern that reflects the net bulk motion of the flow, shaped by the tail itself, and defining a turbulent part  is non trivial; second,  given that the pulsar spin axis belongs to the $z-y$ plane, any memory of the injection condition could translate into a difference in the level of turbulence along the $x$ and $y$ directions; third, by taking planes at different $z$ we can sample how the turbulence varies moving along the tail. Our structure function is defined as:    
\begin{equation}\label{eq:FStruct}
	S_2(\bm{\ell}) = \langle  |  v_{x,y} (\bm{r} +\bm{\ell}) - v_{x,y} (\bm{\ell})  |^2   \rangle\,,
\end{equation} 
where $ \bm{\ell}$ is the displacement vector in the $(x,y)$ plane.
In general, for fully developed incompressible and compressible MHD turbulence, the structure function is shown to scale as $\propto \ell^{\zeta_p}$ ,\citep{She:1994, Grauer:1994, Politano:1995, Muller:2000, Boldyrev:2002, Padoan:2004}, with $\zeta_p= p/9 + 2 -2(2/3)^{p/3}$, such that $S_2(\ell) \propto \ell^{0.7}$.

\begin{figure}
	\centering
	\includegraphics[width=.5\textwidth]{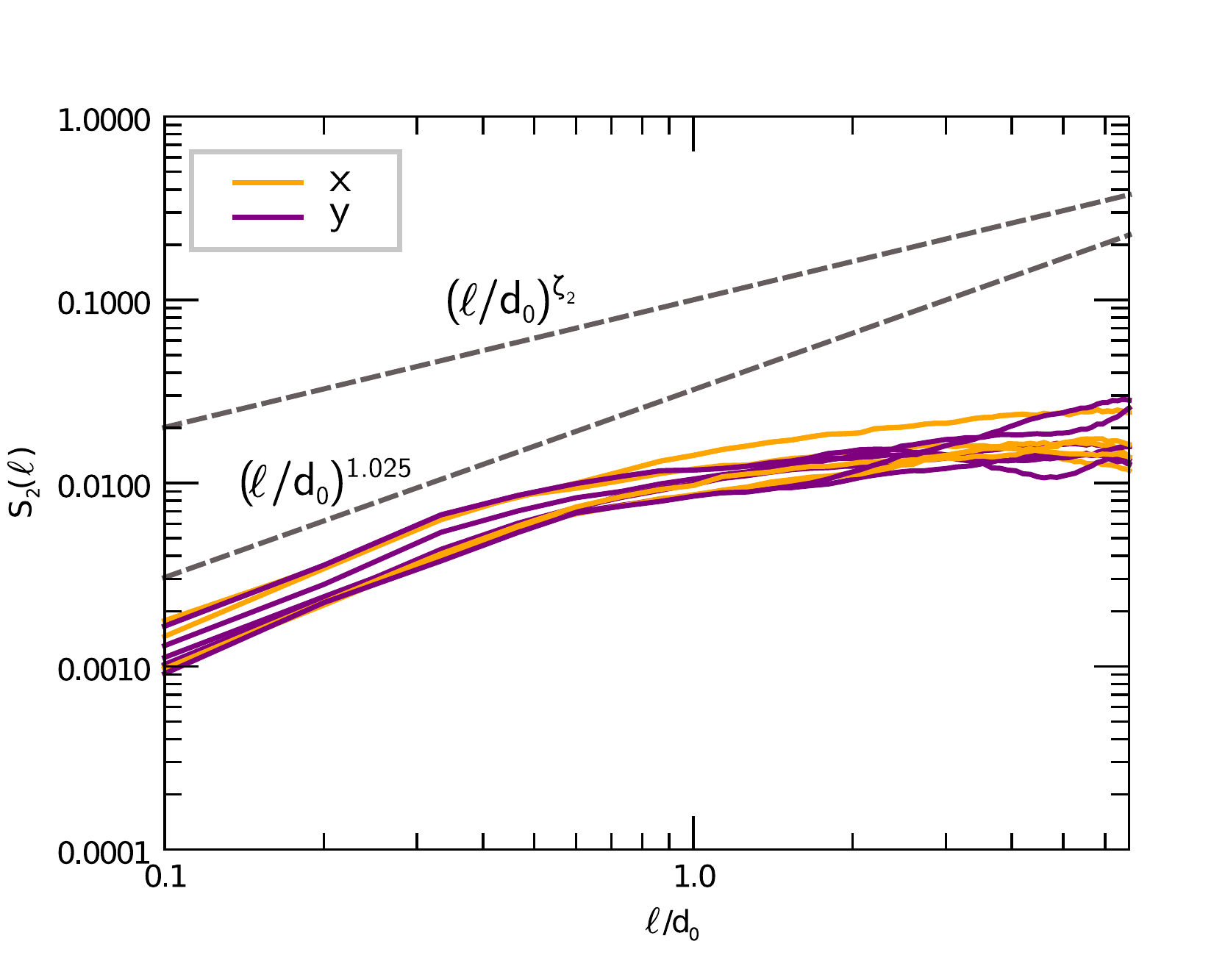}	
	\caption{Second order structure functions for $v_x$ (orange) and $v_y$ (purple) at different cuts along $z$, as functions of the displacement $\ell/d_0$, in log-log scales. The analytical expected power law and the one found in \citet{Zrake:2012} are shown as dashed lines as  comparison.}
\label{fig:FStruct}
\end{figure}

In order to increase out statistics we consider different slices in $z$ in the same range used for Fig.~\ref{fig:turbC} (i.e. for $z\in[-15,\,-7]d_0$) and square sectors in the orthogonal plane with extension [-4, 4]$d_0$ in the $x$ and $y$ directions. The second order structure function is shown in Fig.~\ref{fig:FStruct}.
Note that moving along the tail away from the pulsar, the values of $S_2(\ell)$ drops. This is probably the effect of the sideway expansion of the tail in the orthogonal plane, which obviously increases with distance from the pulsar. We found that our results do not match the analytical prediction: $\zeta_2\simeq 0.7$. This should not be surprising at large scales $\gtrsim d_0$, that are  compatible with the injection scale given by the tail size. On the other hand,  at small scales, we found a trend compatible with the value $\zeta_2=1.025$  found in relativistic MHD simulations of turbulence in low-magnetized plasmas \citep{Zrake:2012}. Scales smaller than $0.1d_o$ cannot be investigated properly due to the limitations of the numerical grid.
We also have verified that these results are not dependent on the resolution or on the AMR resampling.

\section{Conclusions}
\label{sec:conclusion}
In this paper we present a complete description of the dynamics of pulsar bow shock nebulae arising from full 3D relativistic MHD numerical simulations.
In order to discuss how the pulsar parameters influence the tail dynamics we have considered a large set of different configurations, listed in Table \ref{tab:tabRuns}. 
In particular we consider three different initial magnetization, from no-magnetization to highly magnetized, with injected values  $\sigma=[0.0,\, 0.01,\, 0.1,\,1.0]$ and different configurations for the magnetic field geometry, with the angle between the magnetic axis and the pulsar velocity inclined by $0,\, \upi/4$ and $\upi/2$.
Since inhomogeneities in the ISM density are expected to influence primarily the morphology of the forward shock but not very much the dynamics within the contact discontinuity, we only consider here a uniform ISM density, with standard properties.
The pulsar wind is modeled considering both a uniform distribution of the energy flux in the wind (the isotropic model) and a non-uniform distribution, with the maximum of the energy flux in the pulsar equatorial plane (the anisotropic model).
This helps in isolating the influence of the wind properties on the overall morphology of the bow shock.

We first compare our findings with results from a 2D simulation with the same model in the hydrodynamic regime.
Our comparison begin with the average behavior of the physical quantities in the tail in the plane orthogonal to the velocity direction. 
Comparison of 3D runs with 2D HD shows a good agreement in the case of the isotropic wind model, with a quasi perfect overlapping for the intermediate value of the magnetization ($\sigma=0.1$). 
This could appear unexpected, since in principle the major similarity may be expected with the lower magnetized case.
But what we found is that for values of the magnetization $\sigma<0.1$ the dynamics is completely dominated by turbulence on small scales. For greater values on the contrary, injection become dominant.
At the light of these findings is then not surprising that the 2D hydrodynamic case is more compatible with the intermediate case.
The difference in the development of the local turbulence and in the level of mixing of the fluid when moving from 2D to 3D can be better appreciated in Fig.~\ref{fig:Densities}, comparing the first two density maps from the leftmost side, that shows exactly the same model from 2D and 3D simulations.

We also found that in the direction aligned with the pulsar motion ($z-$direction) the velocity shows the same structure obtained in 2D for the isotropic wind model, with  a lower velocity channel around the $z$ axis ($\sim 0.65c$ in 2D and $\sim 0.7-0.8c$ in 3D) surrounded by an higher velocity flow (with $\sim 0.85c$ in 2D and up to $0.9c$ in 3D). 
The same similarity is not seen when comparing 2D HD behavior with the anisotropic wind model. In that case velocity is even maximum along the symmetry axis, with the only exception of run A$_{\{ 0,0\}  }$. 
Moreover the peak velocity is always lower or equal to the maximum value of the 2D HD case.
In the anisotropic cases we also found that the effect of turbulence is even more pronounced: injection becomes important only in for $\sigma>0.1$, even if structures in the flow remain coherent on scales that are much shorter than the same for the corresponding isotropic case. 
The high level of mixing of the fluid clearly shows the effect of the presence of turbulence.

The same considerations remain valid for the magnetic field.
The initial configuration of the magnetic field, and the structured current sheet separating the field polarities, survive in the isotropic cases, while in the anisotropic ones turbulent mixing on small scales tends to destroy the initial configuration, which is only poorly recognizable in the tail, even in the case of higher magnetization.
We found that turbulence does not amplify the magnetic field efficiently: when dynamics is dominated by turbulence rather than injection, magnetic field 
appears to not increase raising the initial value of the magnetization.
On the contrary, as soon as turbulence becomes no more dominant (as it happens for $\sigma \gtrsim 0.01$ in the isotropic model and for $\sigma \gtrsim 0.1$ in the anisotropic one), magnetic field starts to increase with the initial magnetization. 
In 3D magnetic field also appears to be enhanced near to the contact discontinuity, as the effect of an efficient shear instability amplification acting at the CD.

From the morphological point of view the forward shock structure is almost identical from case to case, with the only exception of the anisotropic cases with $\phi_M=\upi/4$, which show the larger deviations from the typical shape. 
Differences in fact appear as small extrusions and blobs, possibly resulting as periodic perturbation of the FS, that will be investigated with major details in following works. Moreover they do not appear to be characterized by high values of the magnetic field and density or pressure, thus it is not clear if they should be visible as emission.
Notice also that they in any case arise on very small spatial scales, possibly making it difficult to be revealed by actual instruments due to resolution limits. 
The observed similarity of the FS between different models should not be surprising, given the forward shock morphology is expected to be primarily influenced by the interaction with the ambient medium and its properties \citep{Romani:1997,Vigelius:2007}, taken as  fixed in our models.
We also confirm that the variation of the magnetization does not reflects in a significant modification of the forward shock, as was previously shown with 2D MHD models \citep{Bucciantini:2005}.

Comparing the bow shocks global morphologies, major differences from case to case are visible mainly in terms of collimation or broadness of the tails, arising as the effect of different magnetic inclination and especially the wind anisotropy. 
These variations of model also influence the global structure and dynamics of the magnetic field, that may be in principle lead to observed differences in the high-energy emission, especially at X-rays, where emission is dominated by the most relativistic particles near to their injection site. This can be less important for radio emission, where the density distribution of emitting particles should be dominant on the magnetic field configuration \citep{Olmi:2014}.
Properties of the emission will be addressed with accuracy in a successive papers, to which we postpone this discussion.

We found that the dynamics in 3D is fully dominated by small scale turbulence for values of the magnetization lower than $\sim 0.1$.This is particularly evident for the anisotropic wind model, where turbulence is more efficient in destroying the injection properties. 
A developed turbulence is well recognizable when looking at the local properties of the flow. 
For the low magnetized anisotropic case with $\phi_M=\upi/4$, which shows the most turbulent dynamics in the tail, we made a more quantitative characterization of the turbulence. We compute velocity structure functions in  a sector of the bow shock tail, considering different slices in $z$. 
We found the velocity structure functions of the two planar components of the velocity, $v_x$ and $v_y$, have a trend which is compatible with results in the literature for MHD relativistic, low-magnetized, turbulence. For scales comparable with the injection scale ($\sim 1 d_0$) we indeed found a flatter trend, that is not surprising since we do expect injection, due likely to shear at the contact discontinuity, to involve different scales along the tail.

Overall our results are in line with to those of \citet{Barkov:2019}. However, due to important differences in the choice of the wind energy and magnetic field distribution, the dynamic of some fluid variables presents some substantial differences. This is not unexpected given that it is well known from standard MHD modeling of PWNe (see for example \citealt{Del-Zanna:2006} vs \citealt{Komissarov:2004}) that these choices affect significantly the post-shock dynamics. For example in our model there is no polar current, which is instead strongly enhanced in the work by \citet{Barkov:2019}, and this impacts not just the distribution of currents in the nebula but, through its dissipation, also the structure of pressure. This can be seen from a direct comparison of our Fig.~\ref{fig:CS} vs their figure 13. 
Moreover we find that for $\phi_M=0$ the thickness of the PWN shocked layer in the head, between the TS and the CD, is much smaller than in their equivalent case. However it can be seen that in their simulations the shock is not properly detached from the boundary of the injection region.

Our results also show that the average flow pattern in the tail is not dissimilar between 2D and 3D runs (major differences are mostly in the head due to inclination and anisotropy). However in the low $\sigma$ regime we found a strongly turbulent magnetic field structure, suggesting that perhaps a laminar model is not likely to fully capture the magnetic field, not even at the level of its average strength. For higher values of $\sigma$ instead, a more coherent magnetic field is found, qualitatively in agreement with the prediction of simplified laminar models \citep{Bucciantini:2017}.
 
We postpone the discussion of the implications of our findings on the observed properties to future works. We in fact intend to present a detailed study of the emission at multi wavelengths, together with the analysis of the time-variability and polarimetric properties of bow shock nebulae from our numerical models, comparing these with available observations. Moreover, it remains to be investigated the escape of  high-energy particles from bow shocks. Since pulsars are known to be one of the most efficient antimatter factories in the Galaxy, the study of the trajectories and escape of particles is of great interest. 
For this problem one must take into account  the external magnetic field and additional configurations need to be investigated depending on the different possible inclinations of the ISM magnetic field with respect to the pulsar velocity. This will be particularly relevant in the far tail of the bow shock, where we found a magnetic field comparable with typical ISM values. 
 
\section*{Acknowledgements}

We acknowledge the ``Accordo Quadro INAF-CINECA (2017)''  for the availability of high performance computing resources and support. Simulations have been performed as part of the class-A project ``Three-dimensional relativistic simulations of bow shock nebulae'' (PI B. Olmi). 
The authors also acknowledge financial support from the PRIN-MIUR project prot. 2015L5EE2Y "Multi-scale simulations of high-energy astrophysical plasmas".
B. Olmi wishes also to acknowledge Andrea Mignone, from the PLUTO team, for fundamental support, Simone Landi and Luca Del Zanna for fruitful discussions.

\footnotesize{
\bibliographystyle{mn2e}
\bibliography{olmi}
}

\end{document}